\documentclass[fleqn,usenatbib]{mnras}

\usepackage[T1]{fontenc}

\DeclareRobustCommand{\VAN}[3]{#2}
\let\VANthebibliography\thebibliography
\def\thebibliography{\DeclareRobustCommand{\VAN}[3]{##3}\VANthebibliography}

\usepackage{graphicx}	
\usepackage{amsmath}	
\usepackage{amssymb}	
\usepackage{newtxtext,newtxmath}

\title[GAMA SOM]{Galaxy And Mass Assembly (GAMA): Self-Organizing Map Application on Nearby Galaxies}

\author[B.W. Holwerda]{Benne W. Holwerda$^{1}$\thanks{Contact e-mail: \href{mailto:benne.holwerda@louisville.edu}{benne.holwerda@louisville.edu}}, 
Dominic Smith$^{1}$,
Lori Porter$^{1}$,
Chris Henry$^{1}$,
Ren Porter-Temple$^{1}$,
\and 
Kyle Cook$^{1}$,
Kevin A. Pimbblet$^{2}$,	
Andrew M. Hopkins$^{3}$,
Maciej Bilicki$^{4}$,
Sebastian Turner$^{5}$,
\and
Viviana Acquaviva$^{6,7}$
Lingyu Wang$^{8,9}$,
Angus H. Wright$^{10}$,
Lee S. Kelvin$^{11}$,
\and
Meiert W. Grootes$^{12}$\\
$^{1}$ University of Louisville, Department of Physics and Astronomy, 102 Natural Science Building, 40292 KY Louisville, USA.\\
$^{2}$ E.A.Milne Centre for Astrophysics, University of Hull, Cottingham Road, Kingston-upon-Hull, HU6 7RX, UK\\
$^{3}$ Australian Astronomical Optics, Macquarie University, 105 Delhi Rd, North Ryde, NSW 2113, Australia\\
$^{4}$Center for Theoretical Physics, Polish Academy of Sciences, al. Lotnik\'ow 32/46, 02-668 Warsaw, Poland\\
$^{5}$Tartu Observatory, University of Tartu, Observatooriumi 1, 61602 Tõravere, Estonia\\
$^{6}$ CUNY NYC College of Technology, 300 Jay Street, Brooklyn NY 11201, USA\\
$^{7}$ Center for Computational Astrophysics, Flatiron Institute, New York, NY, 10010, USA\\
$^{8}$ SRON Netherlands Institute for Space Research, Landleven 12, 9747 AD, Groningen, The Netherlands\\
$^{9}$ Kapteyn Astronomical Institute, University of Groningen, Postbus 800, 9700 AV Groningen, the Netherlands\\
$^{10}$ Ruhr University Bochum, Faculty of Physics and Astronomy, Astronomical Institute (AIRUB), German Centre for Cosmological Lensing, 44780 Bochum, Germany\\
$^{11}$ Department of Astrophysical Sciences, Princeton University, 4 Ivy Lane, Princeton, NJ 08544, USA\\
$^{12}$ Netherlands eScience Center, Science Park 140, 1098 XG Amsterdam, The Netherlands\\
}

\date{Accepted XXX. Received YYY; in original form ZZZ}

\pubyear{2015}

\begin{document}
\label{firstpage}
\pagerange{\pageref{firstpage}--\pageref{lastpage}}
\maketitle

\begin{abstract}
Galaxy populations show bimodality in a variety of properties: stellar mass, colour, specific star-formation rate, size, and S\'ersic index. These parameters are our feature space. 
We use an existing sample of 7556 galaxies from the Galaxy and Mass Assembly (GAMA) survey, represented using five features and the K-means clustering technique, showed that the bimodalities are the manifestation of a more complex population structure, represented by between 2 and 6 clusters.
Here we use Self Organizing Maps (SOM), an unsupervised learning technique which can be used to visualize similarity in a higher dimensional space using a 2D representation, to map these five-dimensional clusters in the feature space onto two-dimensional projections. To further analyze these clusters, using the SOM information, we agree with previous results that the sub-populations found in the feature space can be reasonably mapped onto three or five clusters. 
We explore where the ``green valley" galaxies are mapped onto the SOM, indicating multiple interstitial populations within the green valley population. 
Finally, we use the projection of the SOM to verify whether morphological information provided by GalaxyZoo users, for example, if features are visible, can be mapped onto the SOM-generated map. Voting on whether galaxies are smooth, likely ellipticals, or ``featured" can reasonably be separated but smaller morphological features (bar, spiral arms) can not. SOMs promise to be a useful tool to map and identify instructive sub-populations in multidimensional galaxy survey feature space, provided they are large enough.

\end{abstract}




\section{Introduction}

Quantitative galaxy classification has relied on luminosity, colour, the type of S\'ersic profile \citep{Sersic68} or colour and mass segregation. In such classifications, bimodalities were very often identified in the local Universe \citep[cf.][]{Graham19b}. In colour space, there are two distinct populations; one with blue optical colours and another with red optical colours and higher stellar masses \citep{Baldry06, Willmer06, Ball08, Brammer09}. These populations were dubbed the ``blue cloud'' and the ``red sequence'' respectively \citep{Driver06, Faber07, Taylor15}. This can be translated into a bimodality of specific star-formation (i.e. relative growth of the galaxy) with a ``star-forming galaxy sequence" and a ``quiescent" population
\citep{Noeske07,Wang16d}. 

Similarly, disk and spheroidal galaxies show a bimodal distribution in their S\'ersic \citep{Sersic63} profiles \citep[][Casura in prep.]{Vulcani14, Kennedy15, Kennedy16a, Kennedy16b, Moffett16a, Moffett16b}, moving from pure disk (S\'ersic index n=1) to pure spheroidal (S\'ersic index n=4). By and large, much of the bimodalities seem to correspond to two populations: disk-dominated, star-forming blue galaxies and spheroidal, ``red and dead" quiescent ones. This bimodality is clearest for stellar masses over $10^{10} ~ M_\odot$, while at lower stellar masses the bimodality disappears \citep{Graham06}.  

\cite{Turner19} explored K-means clustering in the multi-dimensional parameter space of nearby galaxies observed with the Galaxy And Mass Assembly \citep[GAMA,][]{Driver09} survey. The high completeness of the spectroscopic redshift component of the GAMA survey combined with multi-wavelength coverage ensured that this is a highly complete census of galaxy populations in the nearby Universe.

\cite{Turner19} found that the bimodalities observed in mass or colour did not translate into direct 
correspondence with morphological features, i.e. not all disk galaxies are blue and star-forming, something already noted by \cite{Masters10a} using GalaxyZoo.
How many actual clusters of galaxy populations there are  remained unclear from the K-means clustering in \cite{Turner19}. More than two and up to six plausible clusters could be identified in the GAMA data using K-means clustering.

Similarly, the notion that the ``green valley" of galaxies in mass-colour space was a single transitioning population has been questioned using galaxy morphology \citep{Schawinski14,Salim15} but it does have more prevalent morphological features 
\citep{Bremer18,Kelvin18} and subpopulations show evidence of quenching \citep{Smethurst15, Smethurst17,Belfiore17, Phillipps19, Bluck20}, though not driven primarily by major mergers \citep{Weigel17}.

There is some evidence that certain morphological subgroups are more common in the green valley \citep[][Smith {\em in prep}]{Bremer18,Kelvin18}, and that major mergers are not very prevalent in the green valley \citep{Weigel17}. However, a sizable sub-population is quenching \citep{Smethurst15,Smethurst17,Rowlands18b,Belfiore17,Phillipps19} and certain morphological features may become more visible as a result, especially given that the quenching appears to be happening inside-out and not galaxy-wide \citep{Bluck20}. 

In this paper, we explore a different unsupervised learning algorithm on the same data-set to visualize the possible sub-populations of galaxies. 
Unsupervised machine learning to explore galaxy morphology is becoming quite common \citep[e.g.][]{Cheng21a,Turner21a}.
We use the Self Organizing Maps \citep{Kohonen} to reduce the dimensionality of the feature data set to a single two-dimensional map. SOM has been used on galaxy morphology and colours before \citep[e.g.,][]{Naim97, Hemmati19, Davidzon19} but the new GAMA feature space opens many new possibilities for application of SOM for galaxy morphology and other properties. 

Our goals are to verify the population clustering underlying the multiple subpopulations that are revealed by the K-means study in \cite{Turner19}, to explore the intermediate population of one bimodality (the green valley population), and evaluate how GalaxyZoo classifications are mapped onto this SOM. 
By mapping the K-means clustering classifications onto a SOM, we will explore how many separate clusters of galaxy populations can be identified in the GAMA data.
Secondly, we will see how the colour bimodality is mapped onto the SOM to identify and map the interstitial green valley population. Lastly, we will map the voting records of the GalaxyZoo project onto the SOM as an alternate label to evaluate how well galaxy-wide feature space can map detailed visual morphology.
Section \ref{s:gama} briefly describes our subsample of the full GAMA catalogs. 
Section \ref{s:kmeans} describes the K-means result and section \ref{s:som} describes the details of SOM training. 
Section \ref{s:results} goes through how the K-means cluster (\S \ref{s:som:kmeans}), green valley (\S\ref{s:som:gv}), and GalaxyZoo votes (\S \ref{s:som:gz}) are mapped onto the SOM. Section \ref{s:concl} contains our concluding remarks.

\begin{figure*}
    \centering
    \includegraphics[width=0.49\textwidth]{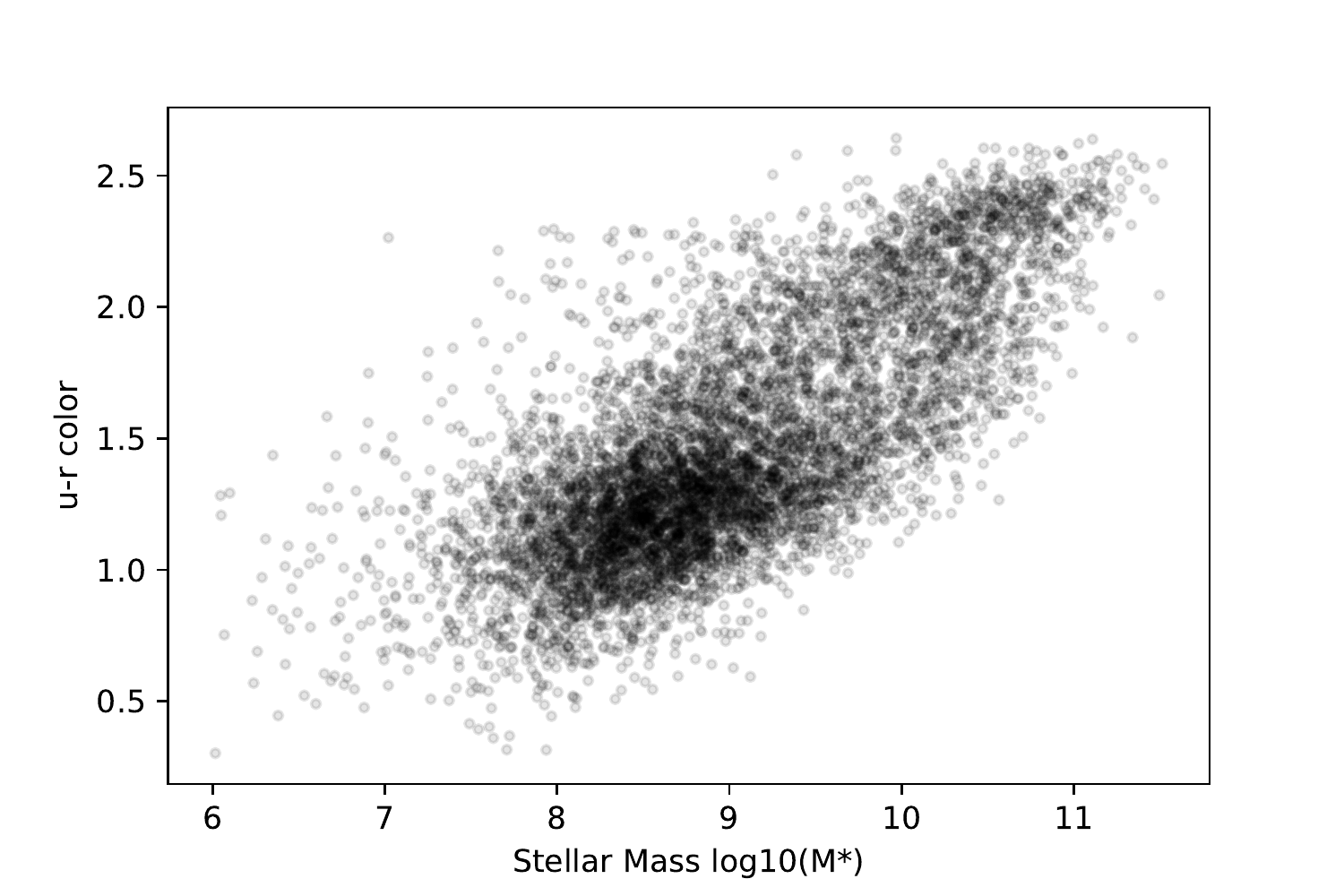}
    \includegraphics[width=0.49\textwidth]{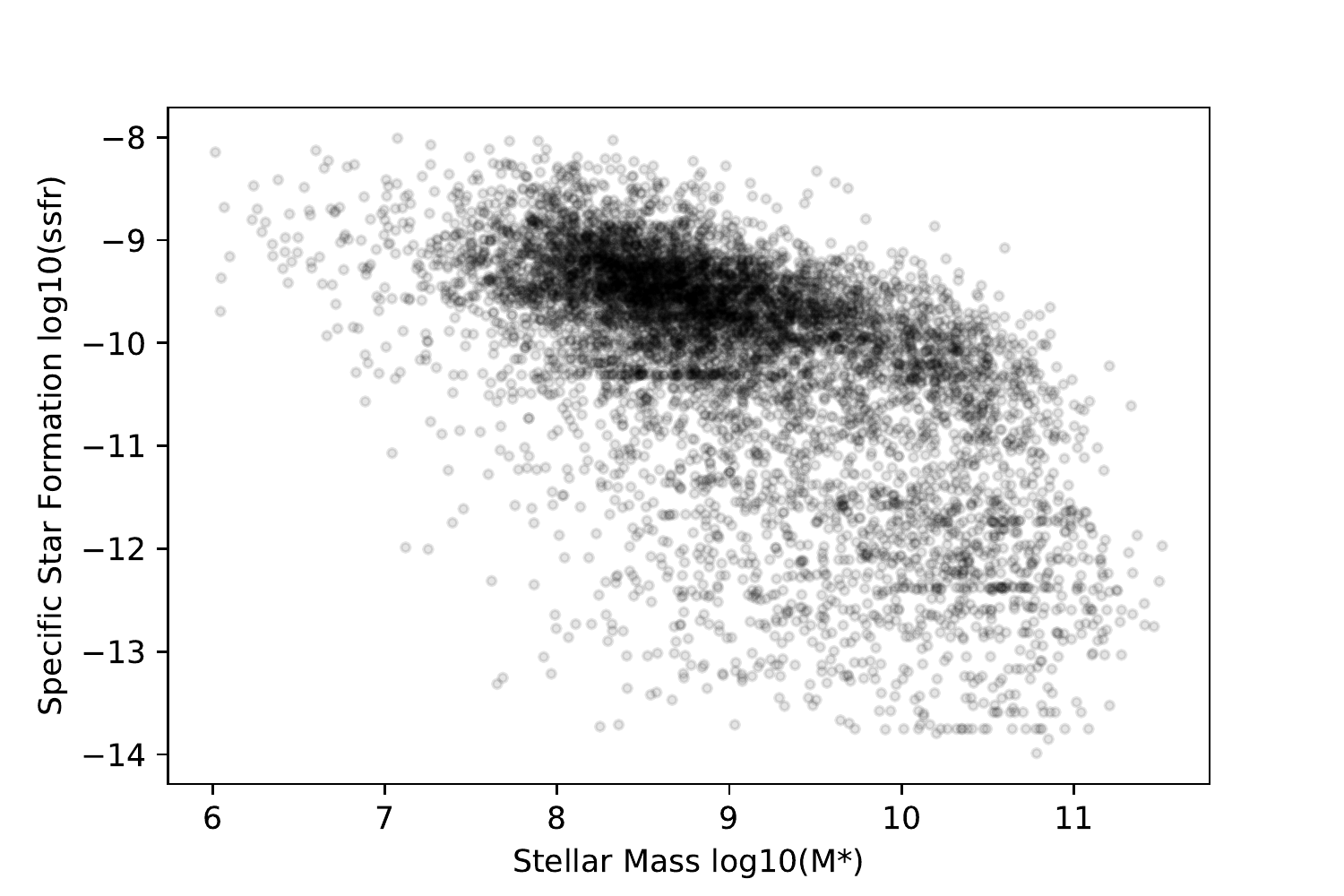}\\
    \includegraphics[width=0.49\textwidth]{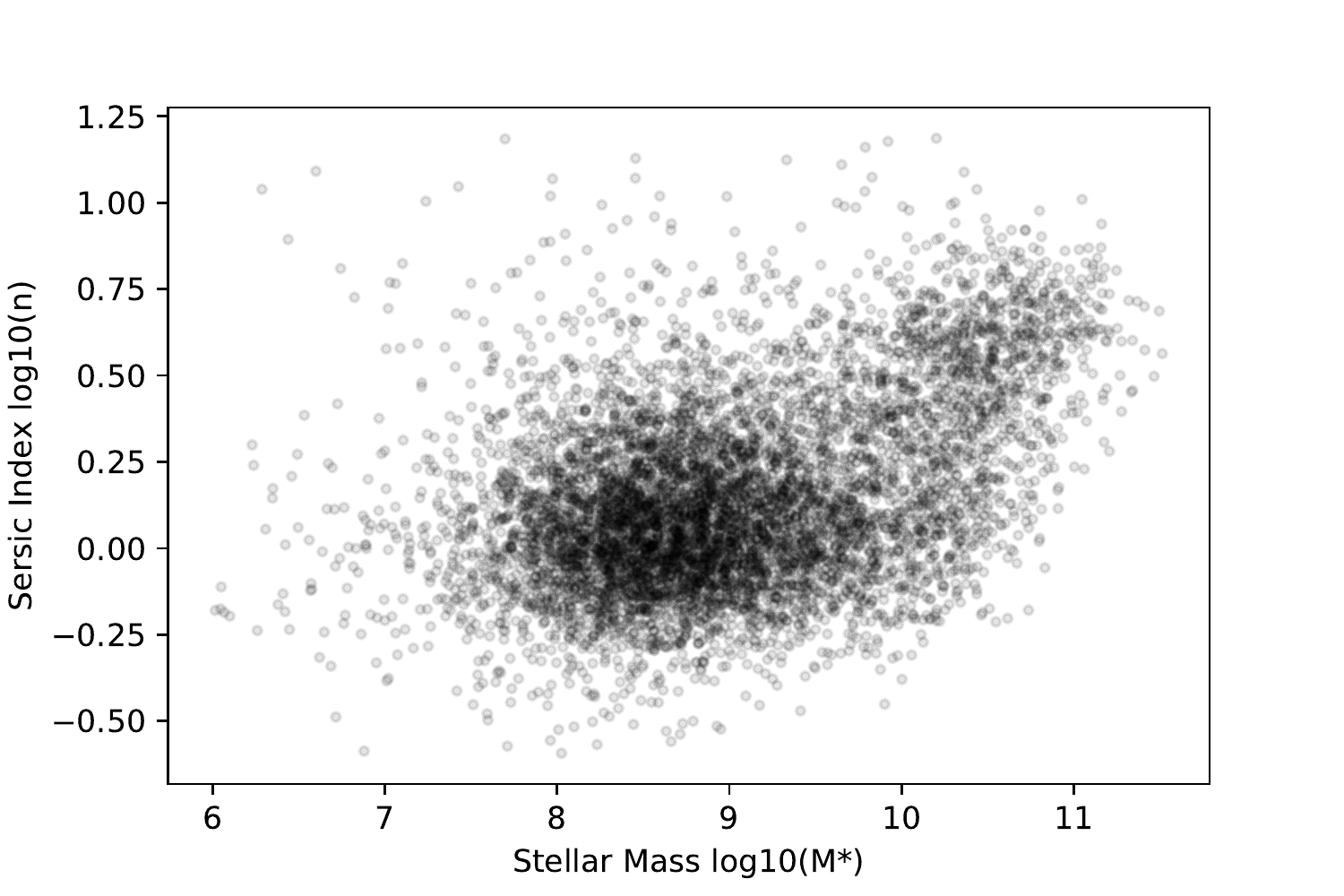}
    \includegraphics[width=0.49\textwidth]{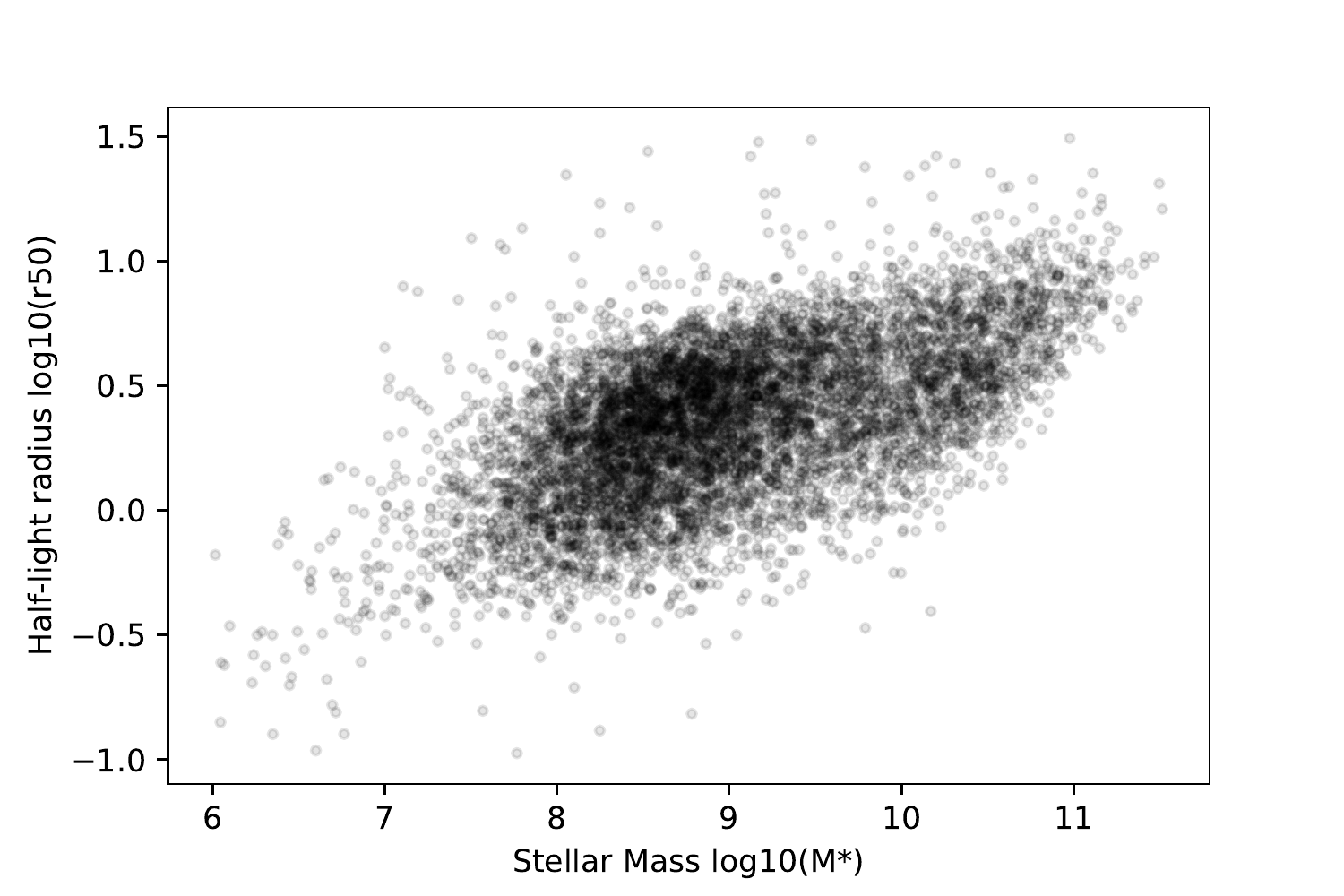}
    \caption{The four features (u-r colour, specific star-formation rate, S\'ersic index $n$, and half-light radius, $r_{50}$) as a function of stellar mass for the GAMA subsample of \protect\cite{Turner19}. Each distribution shows the bimodality often with one strong locus and one less pronounced.}
    \label{f:gama:features}
\end{figure*}

\section{GAMA Subset}
\label{s:gama}

We will use the exact same input dataset derived from the GAMA equatorial catalogs as did \cite{Turner19} for their K-means clustering analysis. 

Briefly, they used redshift and mass limited samples from  phase II of the Galaxy And Mass Assembly (GAMA) survey \citep{Driver09, Liske15}. The main aim of the survey is to study cosmic structure on scales ranging from 1 kpc to 1 Mpc, focusing on groups and their environments. The survey is centered around the spectroscopic campaign, conducted with the
Anglo-Australian Telescope using the AAOmega spectrograph \citep[target catalog in][]{Baldry10}. Reliable redshifts are available for 238000 objects to a limiting r-band magnitude of 19.8 and across five regions covering a total area of 286 deg$^2$. 

The spectroscopic component of GAMA has been supplemented with reprocessed imaging in 21 bands from a variety of other surveys \citep[e.g. the Sloan Digital Sky Survey;][]{York00} and the Kilo-Degree Survey \citep[KiDS][]{de-Jong13,de-Jong15,de-Jong17,Kuijken19} that overlap with the GAMA spectroscopic campaign footprint \citep[the Panchromatic Data Release][]{Driver16}.
Value-added data derived from these spectra and images are listed in tables hosted at \url{http://www.gama-survey.org}.

\cite{Turner19} modeled their sample after that of \cite{Moffett16a}: this is
a low-redshift ($0.002 < z < 0.06$) and magnitude-limited ($r_{PETRO} < 19.8$) sample of 7556 local objects that have been morphologically classified using the method of \cite{Kelvin14}. 

In addition to the \cite{Kelvin14} morphological classification, there are now Galaxy Zoo classifications \citep[][Kelvin et al., \textit{in prep.}]{Lintott08} with votes on disk or spheroid, prominence of bulges, shape of bulges, number and winding of spiral arms, and rarer morphology such as mergers \citep[cf][for dust lanes in this GalaxyZoo data]{Holwerda19}.

The \cite{Turner19} sample's feature space consisted of stellar mass ($M_*$), $u-r$ colour, specific star-formation rate  (ssfr), S\'ersic index ($n$) and half-light  ($r_{50}$). The stellar masses ($M_*$) and the specific star formation (ssfr) on a Gyr timescale are from the MAGPYS \citep{da-Cunha08} 21-filter SED fit catalog (MagPhysv06) described in \cite{Driver16}. The input for this SED fit is the LambdarCatv01 \citep{Wright16}. The restframe $u - r$ colour is based on this LAMBDAR photometry, corrected for redshift ({\sc StellarMassesLambdarv20}), using the same formalism as \cite{Taylor11}, but not corrected for dust effects.
The r-band S\'ersic indices (n) and half-light radii ($r_{50}$) are from the analysis of Sloan Digital Sky Survey images described in \cite{Kelvin12}, {\sc SersicCatSDSSv09} in GAMA repository. These catalogs are readily available at \url{http://www.gama-survey.org/dr3/}.

Our feature selection is entirely based on the one from \cite{Turner19} because it is in this space is where the bimodalities occur. 
The feature space is a mixture of observed (e.g. restframe $u-r$ colour) and derived parameters (e.g. specific star-formation rate). The derived values may have systematic effects in their values (e.g. mass-to-light ratio for stellar masses). However, the feature space is whitened (scaled to standard deviation and mean set to 0) and should suffice for relative discrimination. 
These are the features on which the K-means clustering was based and we will train our Self Organizing Map on. 

We should note here that the feature sample is not evenly balanced between bimodalities. Figure \ref{f:gama:features} shows the distributions of the features used in the K-means clustering. Each panel shows signs of the bimodalities noted earlier but like most real data, the clusters are not neatly separated and not with equal representation in the data.

\begin{figure*}
    \centering
    \includegraphics[width=0.49\textwidth]{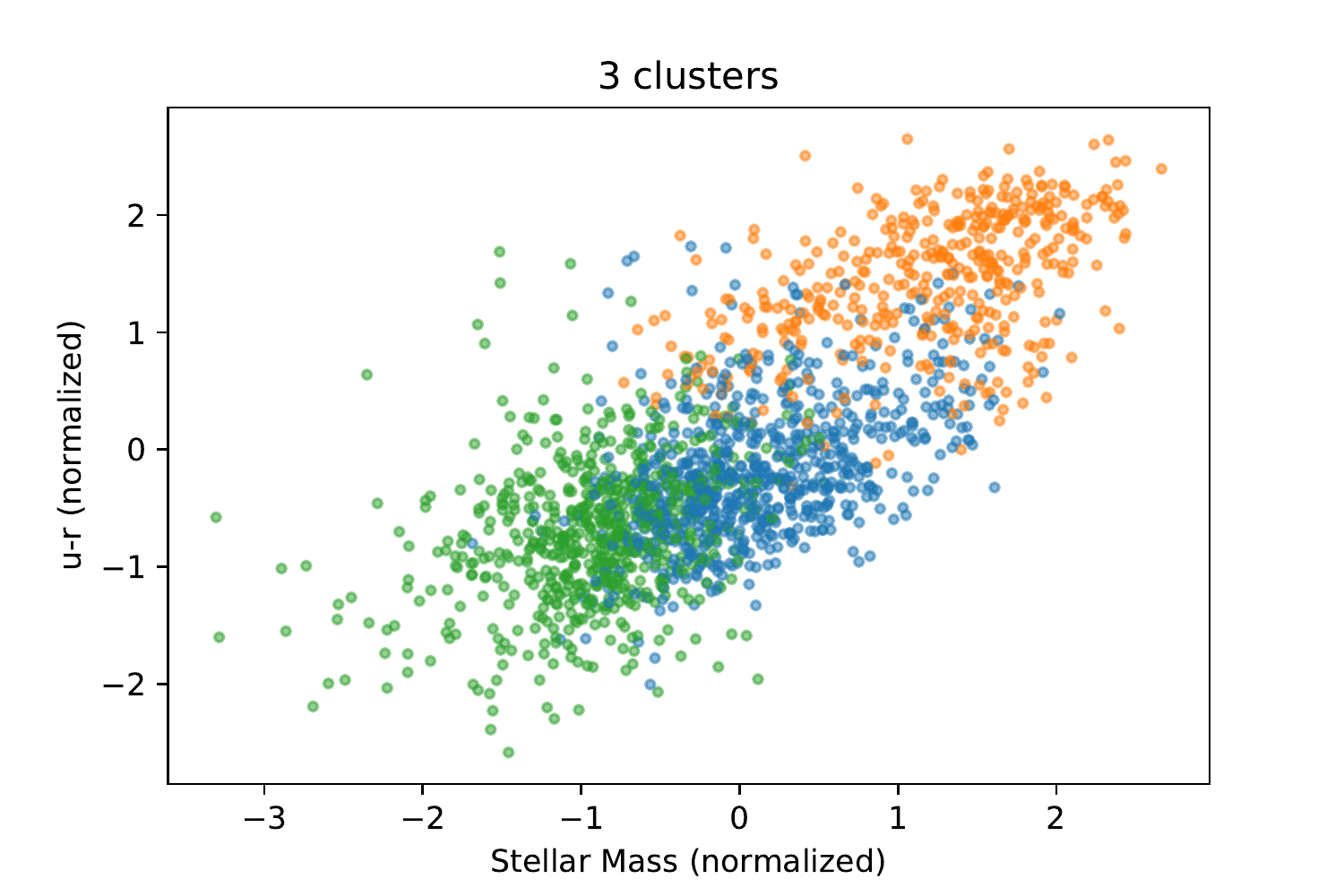}
    \includegraphics[width=0.49\textwidth]{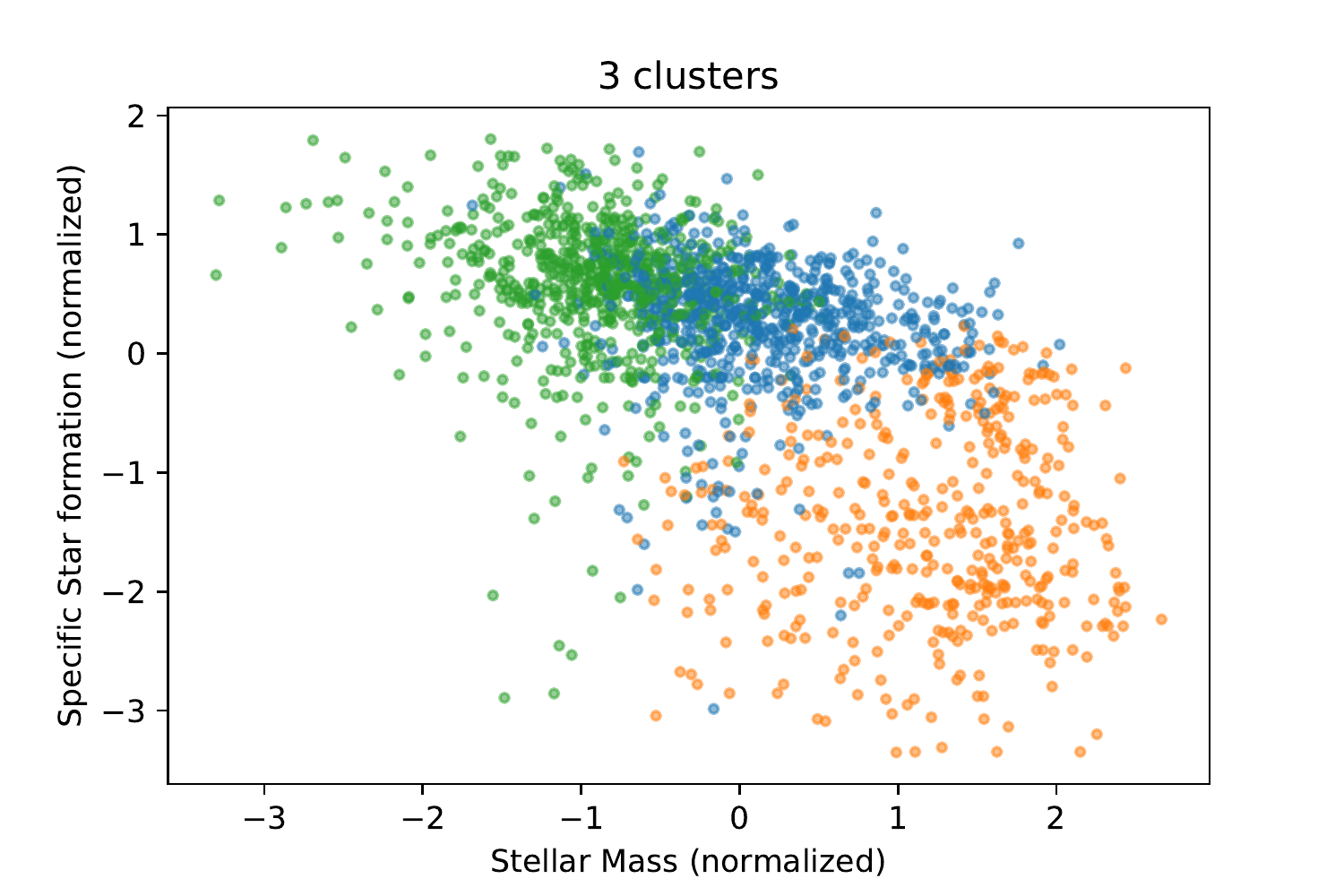}\\
    \includegraphics[width=0.49\textwidth]{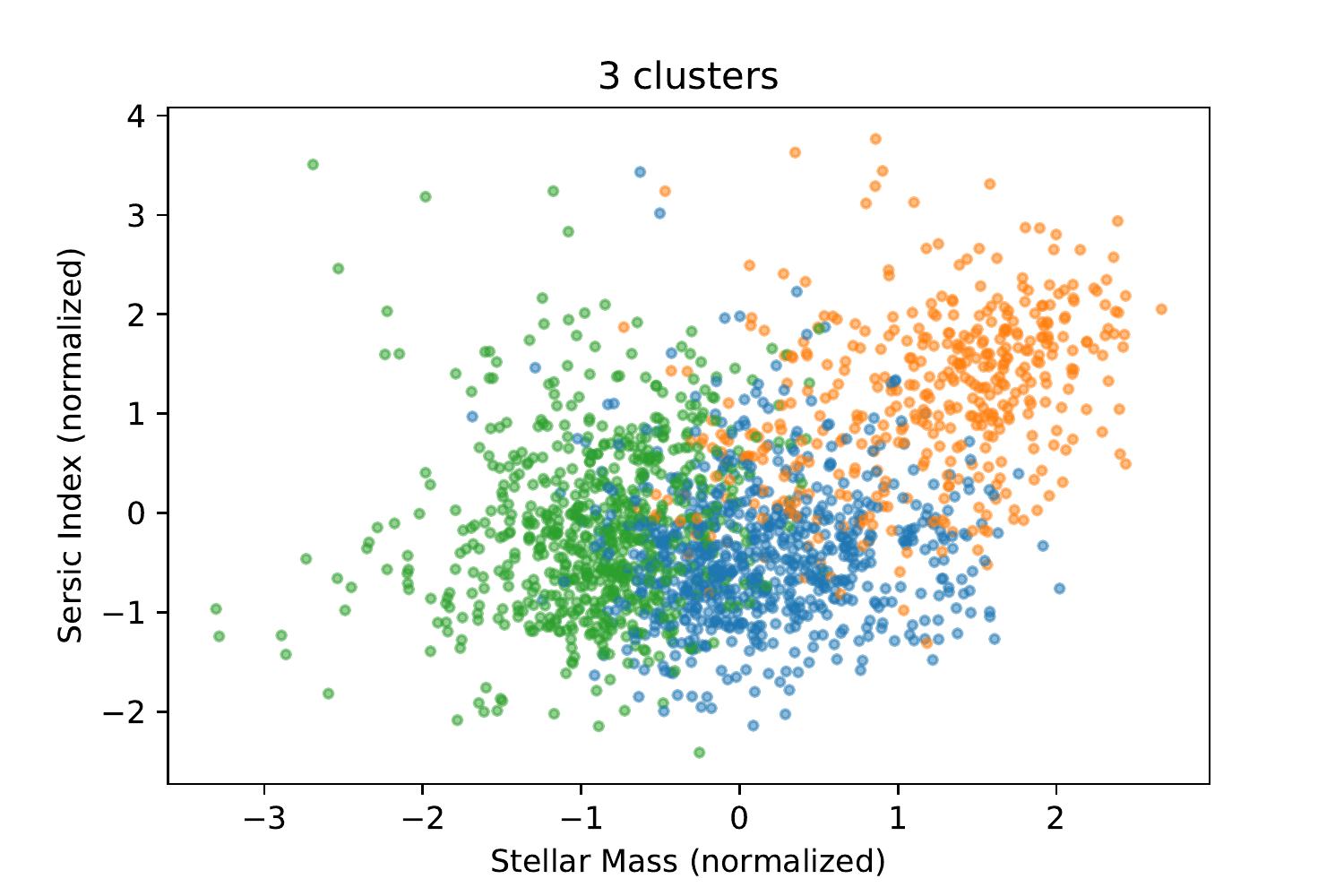}
    \includegraphics[width=0.49\textwidth]{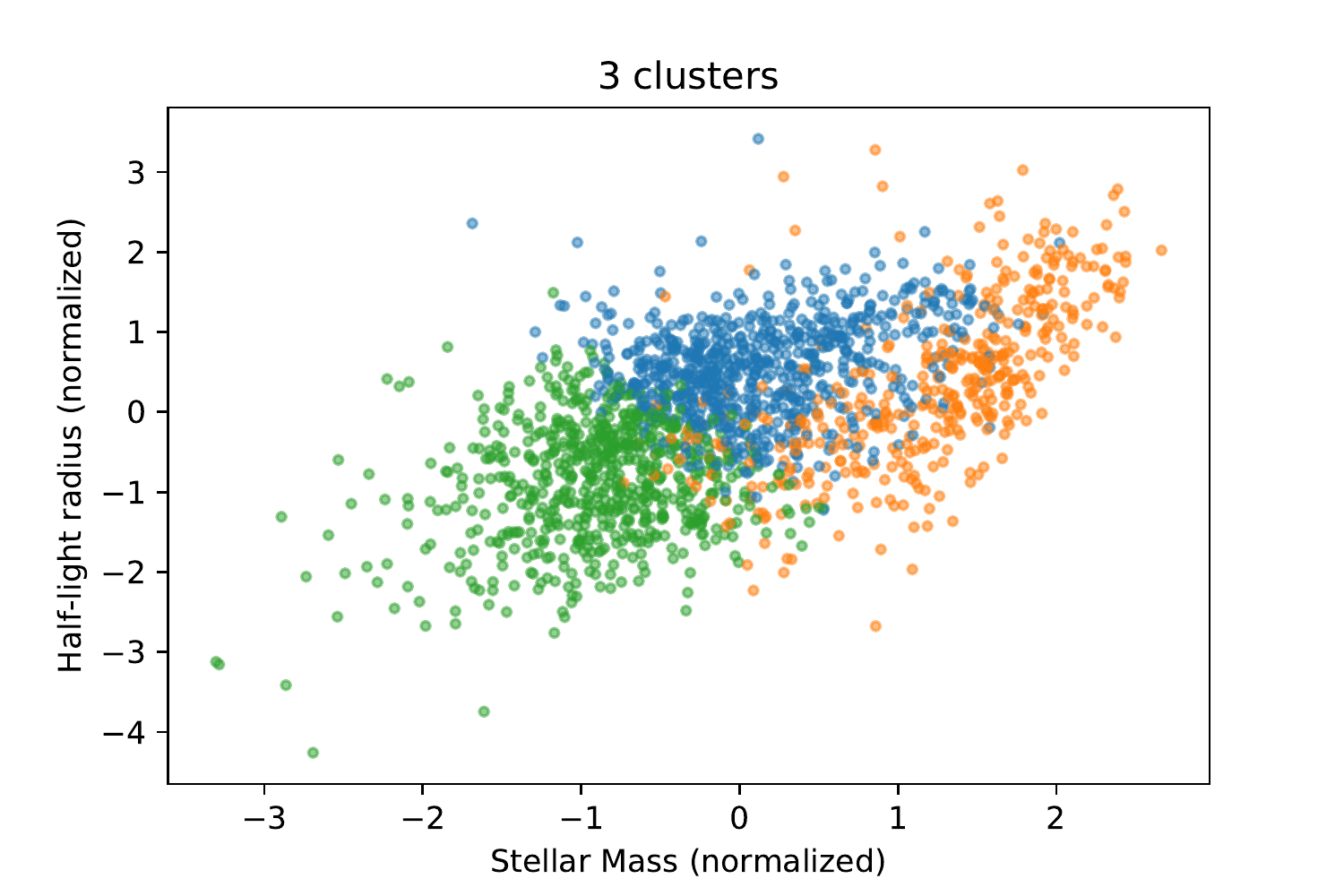}
    \caption{The four features (u-r colour, specific star-formation, S\'ersic index n and half-light radius $r_{50}$) as a function of stellar mass, now all normalized and whitened for use by the K-means clustering algorithm. The 3 K-means clusters from \protect\cite{Turner19} identified are indicated in the normalized feature space.}
    \label{f:k3:features}
\end{figure*}

\begin{figure*}
    \centering
    \includegraphics[width=0.49\textwidth]{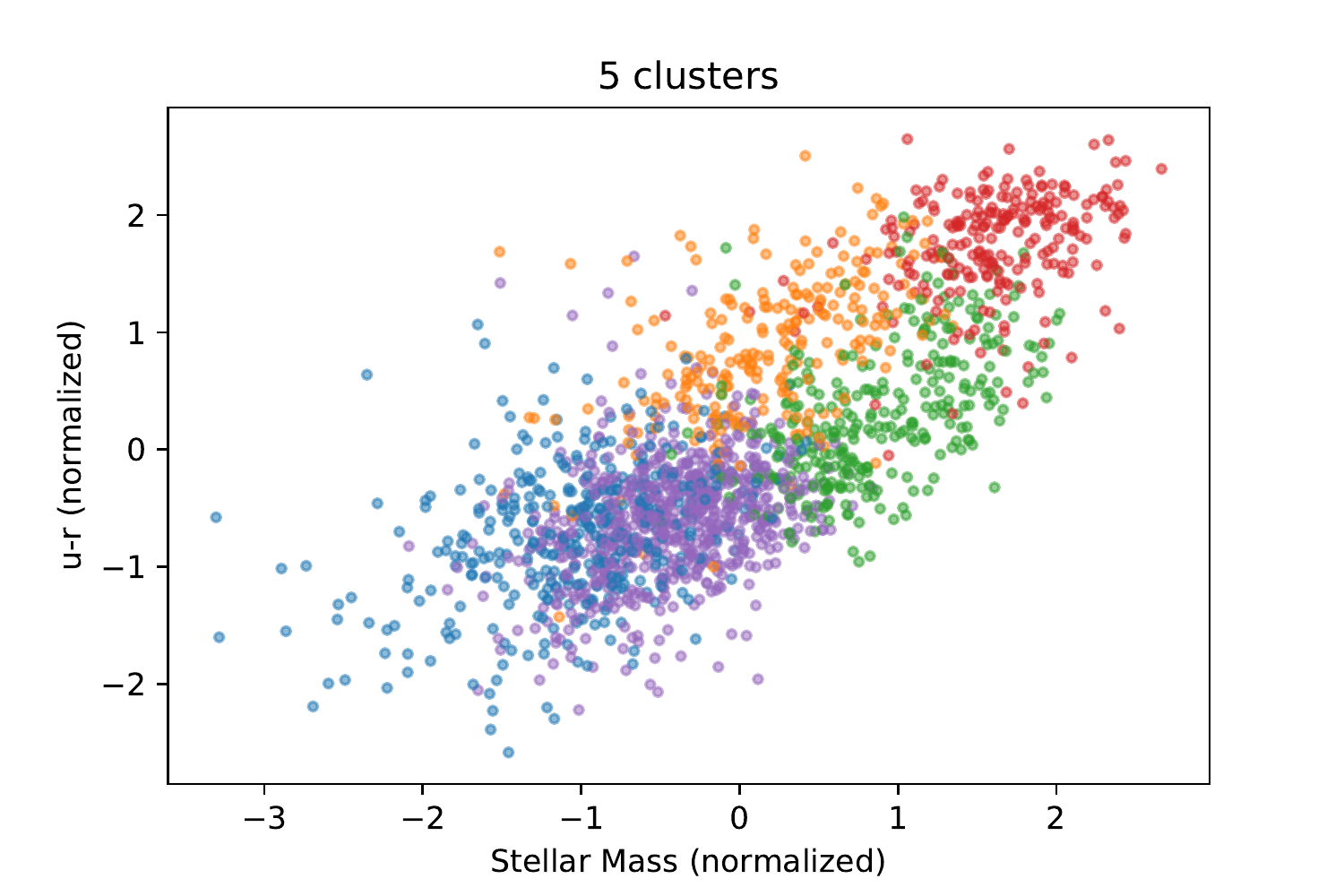}
    \includegraphics[width=0.49\textwidth]{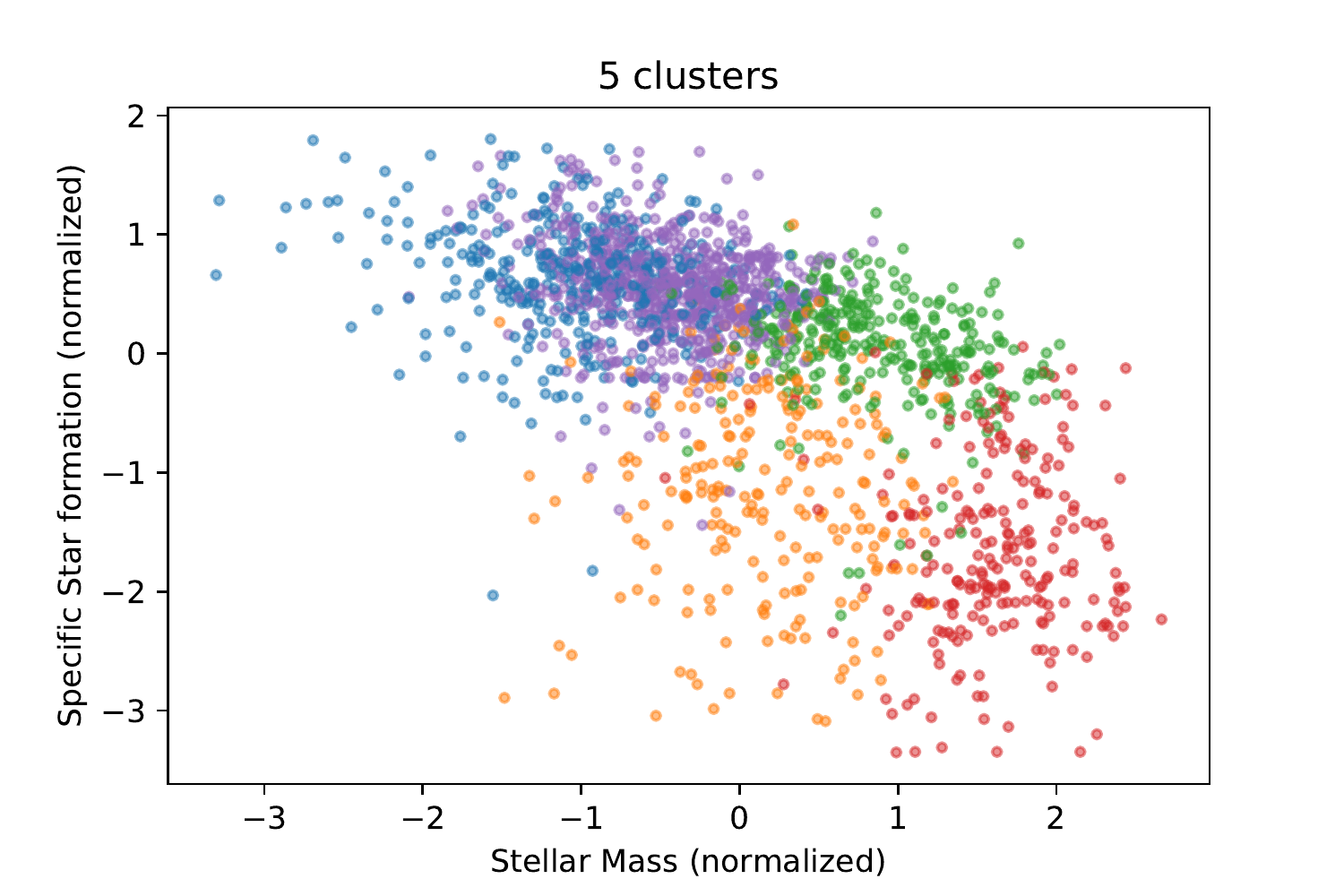}\\
    \includegraphics[width=0.49\textwidth]{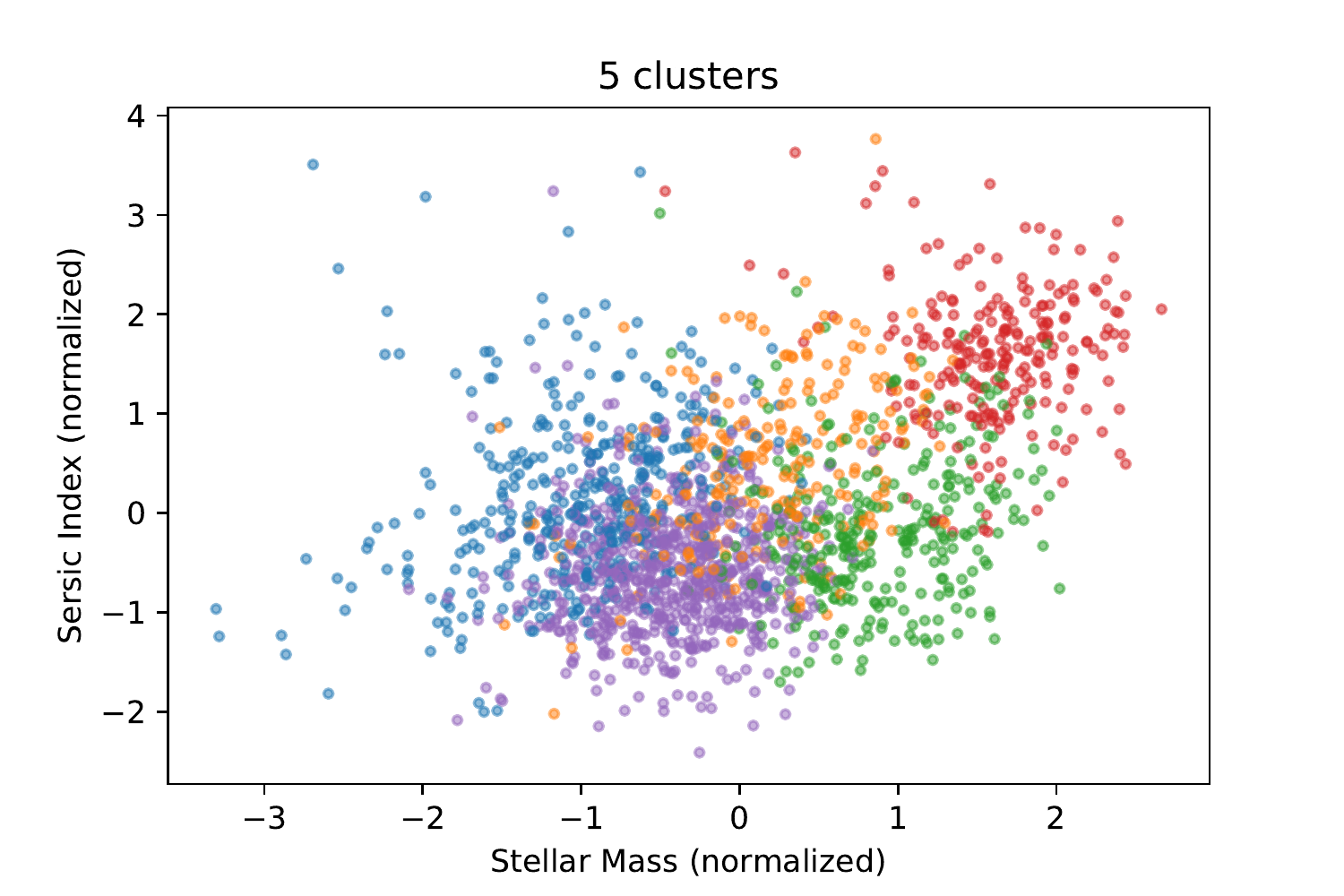}
    \includegraphics[width=0.49\textwidth]{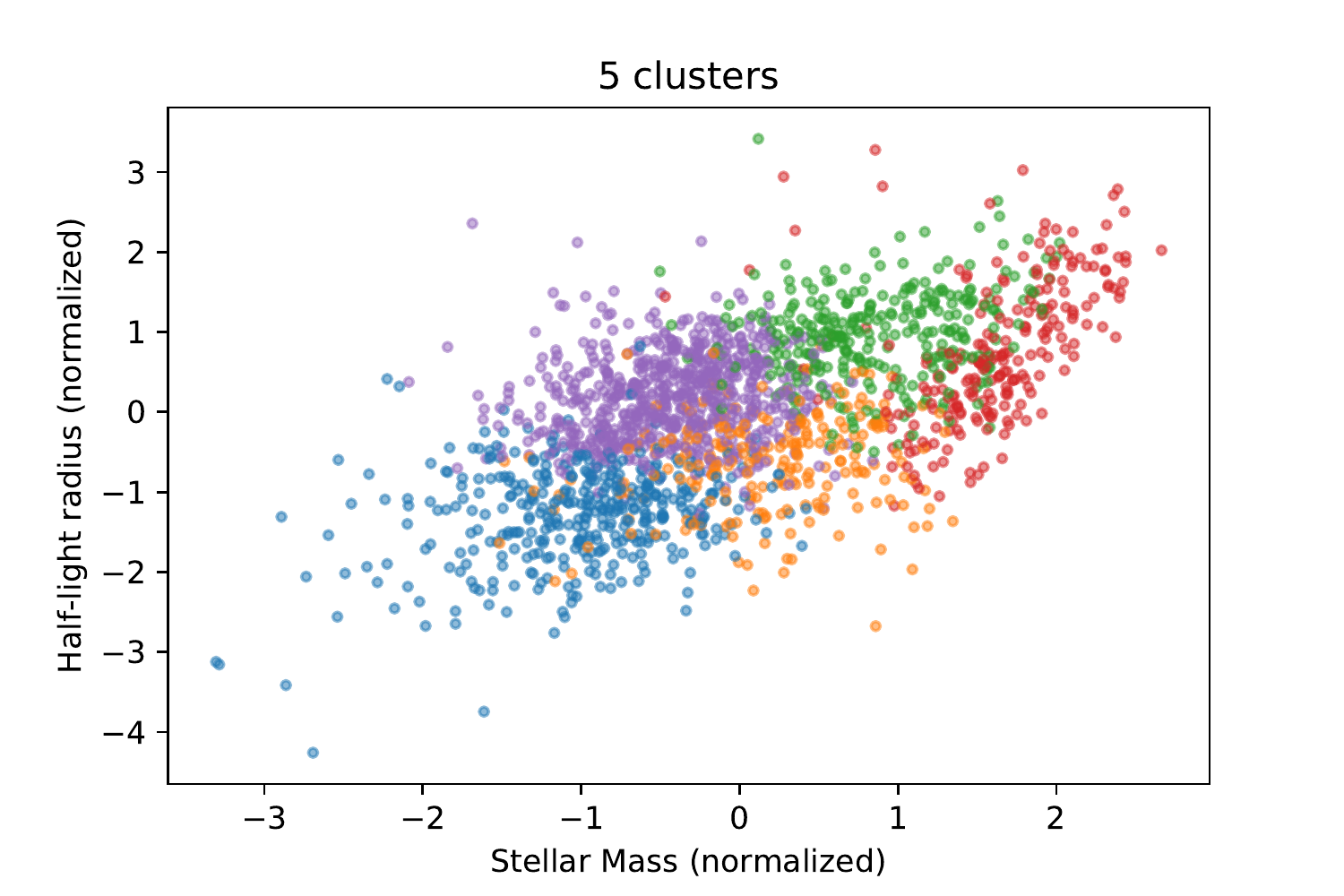}
    \caption{Same as Figure \ref{f:k3:features} but now for the 5 cluster space indicated with colours.}
    \label{f:k5:features}
\end{figure*}

\section{K-means Clustering}
\label{s:kmeans}

Although K-means worked well on this data-set, it is important to reiterate that a drawback of K-means method is that we have to specify the number of clusters, $k$, as an input. The optimal value of $k$ is generally not known in advance, hence the optimal value must be identified by clustering at several different values of $k$ and analysing the outcomes afterwards. The number of clusters\footnote{This is the K-means cluster method's hyperparameter.} to choose is not necessarily obvious in real-world applications such as these, especially in the case of a higher dimensional data-set or one that is not distributed equally among the number of clusters. This is an issue \cite{Turner19} encountered and noted as well. 
Figures \ref{f:k3:features} and \ref{f:k5:features} illustrate the clustering in the whitened feature space. Both a 3-cluster or 5-cluster solution is just as visually appropriate. 

We will now first examine the optimal number of clusters in this sample, using the K-means clustering implementation in {\sc scikit-learn}. In \cite{Turner19}, the optimal number of clusters was addressed by examining the stability of clustering outcomes over large numbers of randomised initialisations. The elbow method is a useful graphical tool to estimate the optimal number of clusters $k$. If $k$ increases, the within-cluster Sum of Squared Errors (SSE, or ``inertia'' or ``distortion'') should decrease, because the samples will be closer to the centroids they are assigned to.

\begin{figure}
    \centering
    \includegraphics[width=0.5\textwidth]{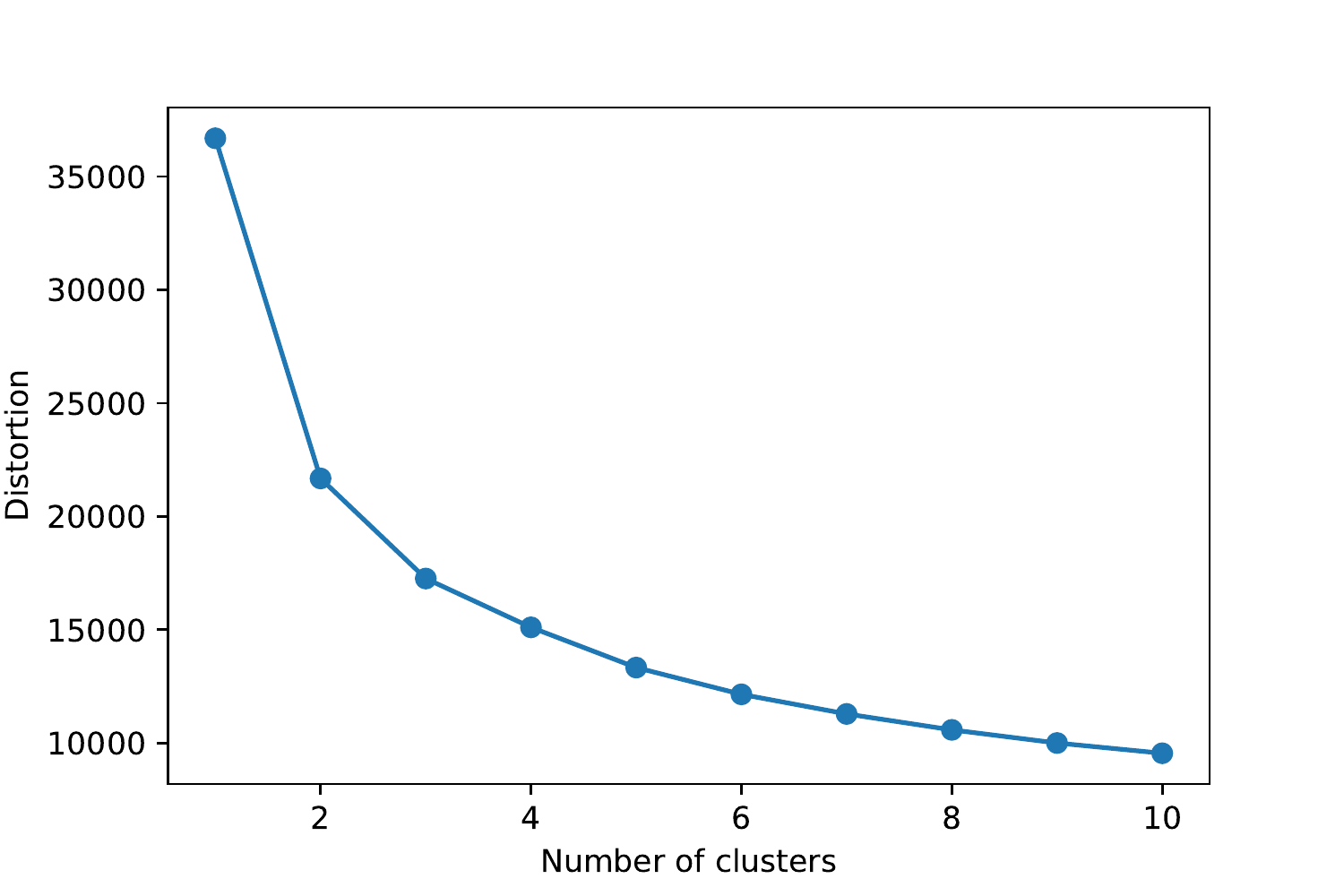}
    \caption{The distortion (also known as inertia or SSE) as a function of the number of clusters employed on the normalized GAMA sample. The conclusion that the optimal number lies somewhere between k=2 and k=5.}
    \label{f:kclusters:optimization}
\end{figure}

The idea behind the elbow method is to identify the value of $k$ where the distortion decreases most rapidly. This is shown in Figure \ref{f:kclusters:optimization} which plots the SSE as a function of number of clusters $k$ for the GAMA data-set. 

We conclude from this plot that the optimal number of clusters is between 2 and 5 with which to classify the GAMA data-set, in agreement with \cite{Turner19}. Three clusters appear optimal, showing that the bimodalities noted in this feature space is not a single underlying bimodality but are comprised of subpopulations.

\begin{table}
    \centering
    \begin{tabular}{l|l l}
    $N_{clusters}$ & $S_{train}$& $S_{test}$\\
\hline
\hline
        2 & 0.395 & 0.407 \\ 
        3 & 0.260 & 0.266 \\ 
        4 & 0.231 & 0.239 \\ 
        5 & 0.238 & 0.245 \\ 
        6 & 0.230 & 0.228 \\ 
\hline
    \end{tabular}
    \caption{The Silhouette coefficients for the K-means classifications for each of the numbers of clusters. Values that are significantly lower than 1, like those observed here, indicate significant overlapping among clusters. }
    \label{t:kmeans:silouette}
\end{table}

To evaluate the K-means clustering here further, we employ the silhouette coefficient ({\sc sklearn.metrics}). The values for the training and test sample for each number of clusters are listed in Table \ref{t:kmeans:silouette}. 
The silhouette coefficient is calculated from the mean intra-cluster distance (a) and the mean distance to the nearest-cluster to which the sample does not belong (b):
\begin{equation}
   S = \langle  (b - a) / max(a, b) \rangle.
\end{equation}
over the whole sample. 
Optimal coefficient value is 1 and the poorest clustering is denoted by -1 (misclassification). Values near 0 indicate overlapping clusters. In an optimal application scheme, clusters are well separated, i.e.  $b >> a$ and $max(a,b) = b$, so the silhouette coefficient is close to 1. Lower values indicate overlapping clusters.

The values in Table \ref{t:kmeans:silouette} show values closer to 0 than 1, indicating that the clusters are not well-separated  (as can be seen in Figures \ref{f:k3:features} and \ref{f:k5:features}). K-means clustering is ideal for isomorphic (rounded) and balanced clusters (approximately equal numbers of objects in each cluster). This may not necessarily be the case here (Figure \ref{f:gama:features}). 
We checked with scikit-learn's Density-Based Spatial Clustering of Applications with Noise (DBSCAN) as an alternate clustering algorithm as well but this performed poorly in comparison (lower or even negative silhouette coefficients). We conclude that in this feature space, there is significant overlap among clusters. 

Self Organizing Maps can be used to generate a 2-dimensional map of a higher-dimensional feature space; vicinity in the SOM space is associated with similarity in the original feature space. This is not the only alternative, for example, one could employ a Principal Component Analysis \citep[PCA][]{Conselice06,Scarlata07} or alternate clustering algorithms \citep{Turner21a}. We opt for SOM for the ease of visualization. Another advantage is that SOMs are non-linear, which helps to preserve both global and local structures from the high-dimensional feature space in the final projection.

\section{Self Organizing Maps}
\label{s:som}

Using the full 7556 sources in the GAMA nearby galaxy sample, we train a $100\times100$ size SOM\footnote{This is an order of magnitude more nodes than the rule of thumb $N_{nodes} \sim 5 \times \sqrt{N}$, where N is the number of data-points} for  1000 iterations using the {\sc minisom} \citep{minisom} Python implementation of Self-Organizing Maps. Input data is the set of features, a vector for each object of Stellar mass, specific star-formation rate, u-r colour, S\'ersic index, and effective radius, all renormalized (also known as ``whitened'') to ensure a mean of 0 and scaling to their standard deviation. 

Self Organizing Maps are an early form of unsupervised learning \citep[cf.][]{Kohonen}. Briefly, a map is seeded with random vector nodes that resemble the (whitened) data in mean and standard deviation. 
During training, a data point is picked at random, and its closest node in the SOM, known as Best Matching Unit or BMU, is found, and the coordinates of the BMU and its surrounding nodes are updated to move towards that data point. At each learning iteration, the full data-set is applied to the SOM, modifying its ``winning'' nodes and neighbors to resemble the data instance more. 
The process is repeated for many iterations; once converged, the final network of nodes of the SOM should be topologically close to the full data set.
A SOM is ideal if no clear number of classes is known beforehand and one wants to map a multi-dimensional feature space onto a single two-dimensional map. 

Because the initialization of the SOM is a randomized event, the learning process for each time the SOM is trained will result in a different SOM. The only hyperparameter for the SOM is its size, which determines its resolving power, and the choice of  learning function, which optimizes how long it takes to converge. 

\begin{figure*}
    \centering
    \includegraphics[width=\textwidth]{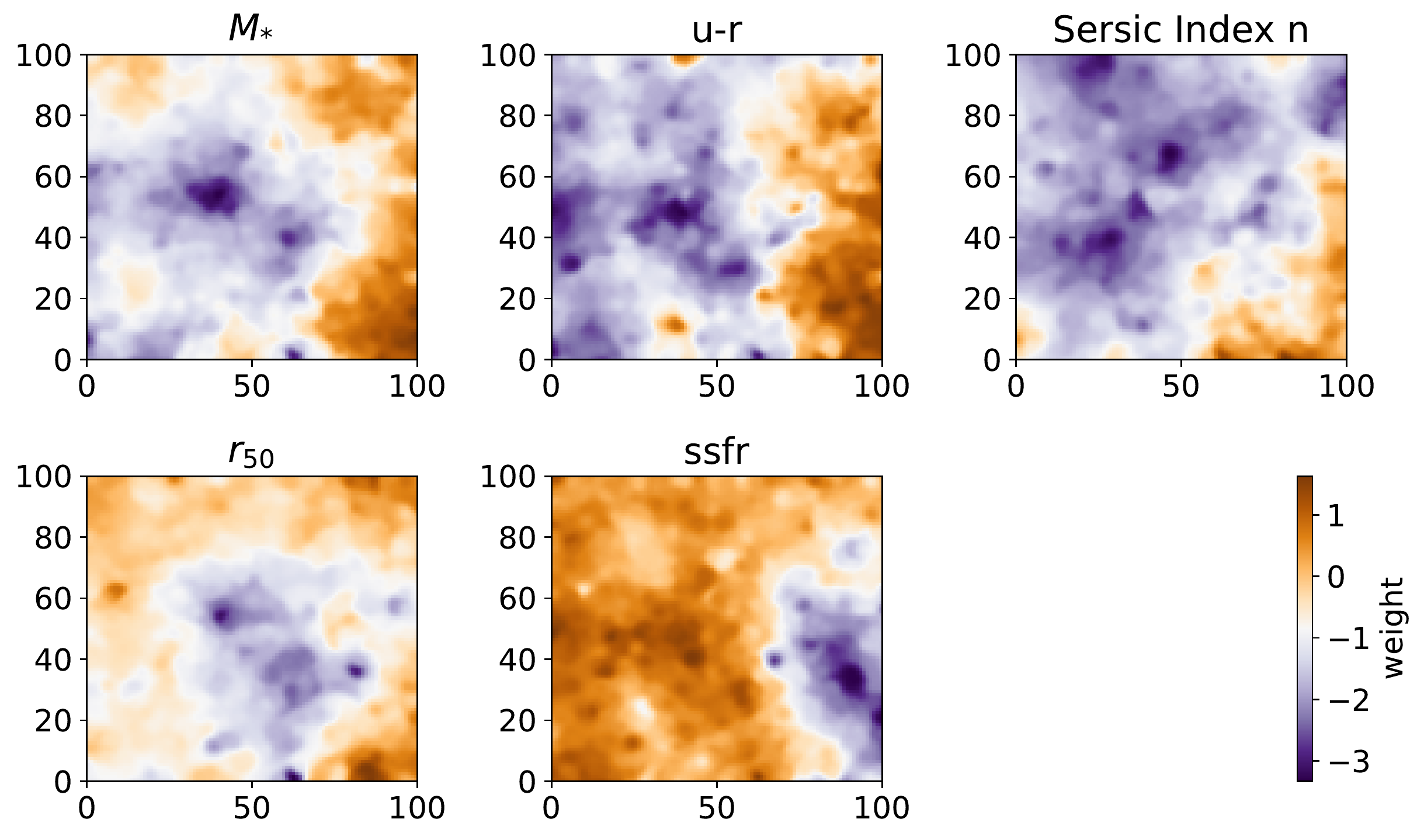}
    \caption{The five features (u-r colour, specific star-formation, S\'ersic index n, half-light radius $r_{50}$ and stellar mass) mapped from the multidimensional parameter space onto our SOM to show their relative weight in each SOM node on the map. }
    \label{f:som:featuremap}
\end{figure*}

The SOM we have generated provides a 2-D representation of our sample where similarity is preserved: objects that are close in the higher-dimensional feature space should be close also in the SOM. Therefore, we can use the SOM as a ``canvas'' onto which we map other properties, such as individual features, or the clustering structure identified by the k-means algorithm, to understand how these relate to the higher-dimensional representation of our sample.
Figure \ref{f:som:featuremap} shows where the feature space is mapped onto this instance of the SOM. 
Because a SOM is initiated with a random seed, the map would appear different after each initiation and training\footnote{In the accompanying notebook, the SOM has been pickled and can be loaded. Or one can opt to retrain.}.
For a SOM different initialisations would still be expected to yield the same results at a qualitative level (i.e. clusters would still appear grouped together, even if they show up somewhere else in the map). We retrained this SOM several times\footnote{Each student in the 2021 Spring P650 class at the University of Louisville ran a version.} each time arriving at the same qualitative result. 
\begin{figure*}
    \centering
    \includegraphics[width=0.49\textwidth]{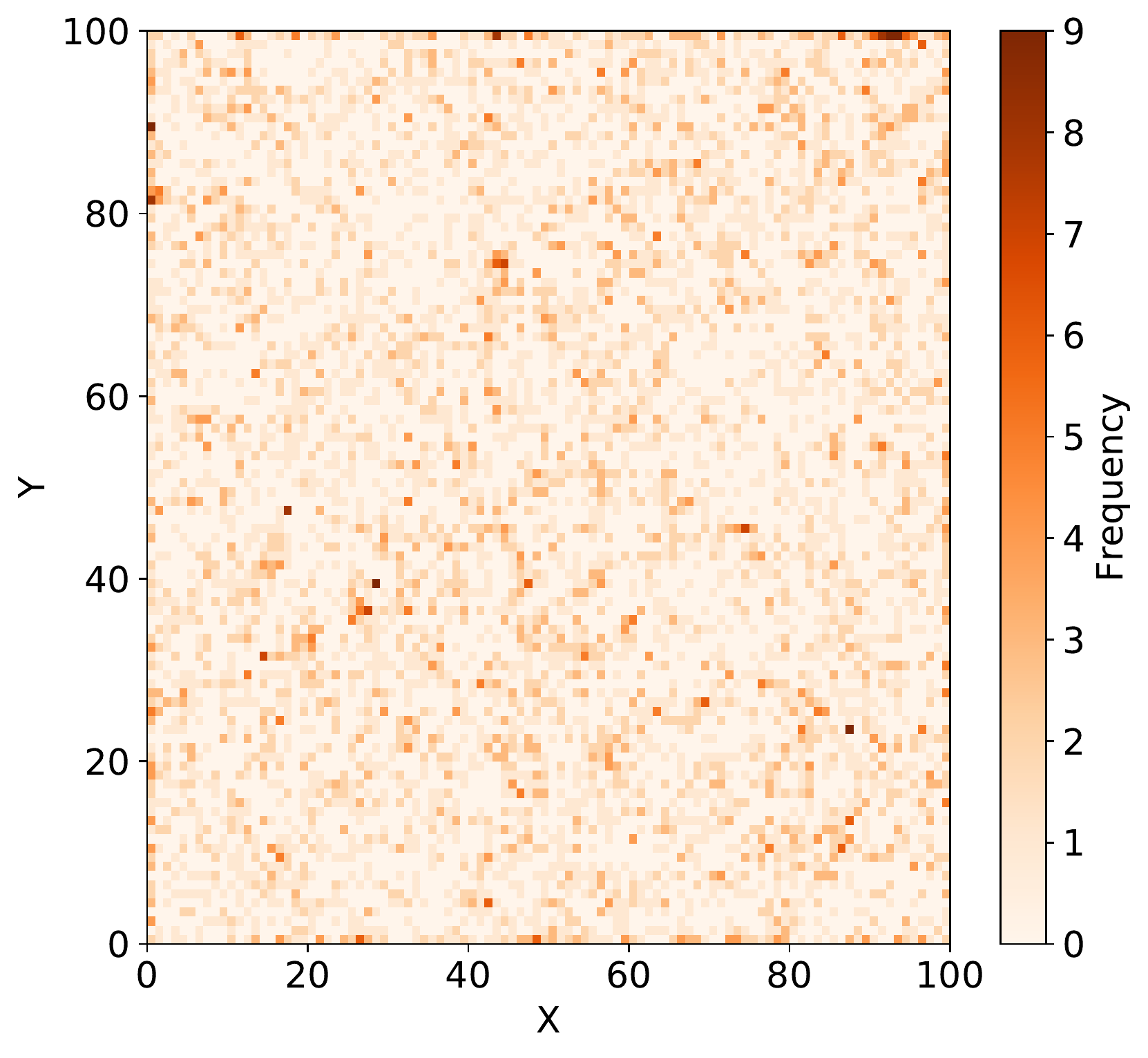}
    \includegraphics[width=0.49\textwidth]{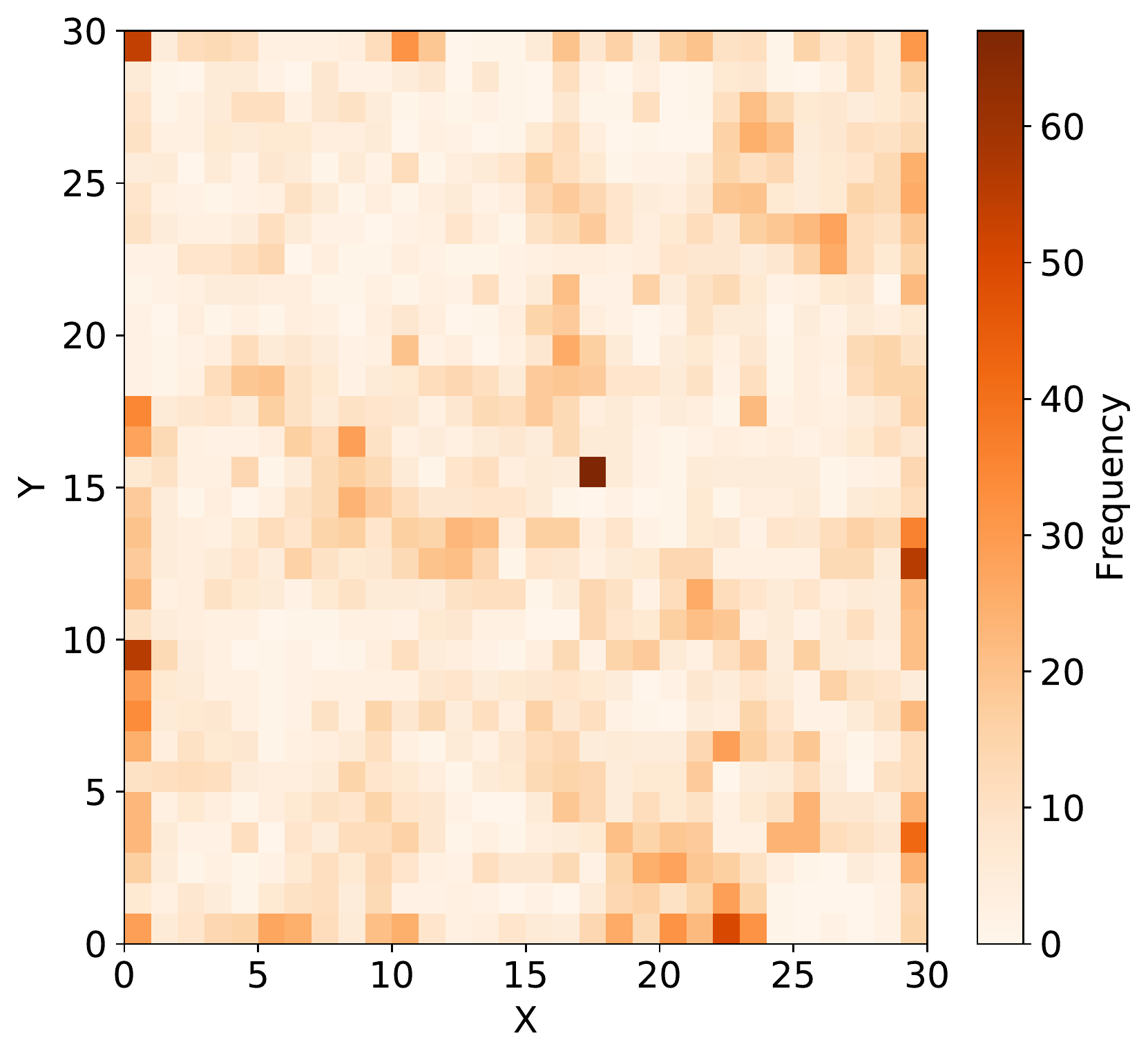}
    \caption{The frequency map of our SOM: the frequency each node is the winning one in classifying the sample. The $100\times100$ (left) and a $30\times30$ (right). The scale of the SOM is the predominant hyper-parameter for SOM. In a lower resolution version, a few nodes attract high numbers of sources, losing resolution in those areas while SOM in-between these remains much lower populated. }
    \label{f:som:frequency}
\end{figure*}

Specific star-formation and $u-r$ colour are most closely related in the weighting of the SOM. This is consistent over multiple training runs of the SOM. This should not surprise us as u-r is a star-formation tracer. It serves to remind us that the feature space is not orthogonal but somewhat degenerate. 

Figure \ref{f:som:frequency} shows the winning frequency of each node when classifying the full sample. Maximum frequency is 9. The overall size of the SOM and resulting total number of nodes may be somewhat generous for the size of this sample.  
Figure \ref{f:som:frequency} compares the frequency of matches between our data and a SOM node for the 100$\times$100 SOM and a 30$\times$30 one trained on the same data-set. The smaller SOM trains faster but corrals objects in a few nodes, leaving less resolution to differentiate sub-populations. 


To evaluate the quality of a feature map, we use two indicators: learning quality and projection quality. Quantization error and topographical error are our main measurements to assess the quality of SOM. Quantization error is the average difference of the input samples ($x(t)$) compared to its corresponding winning map point ($w(t)$). It assesses the accuracy of the represented data, therefore, it is better when the value is smaller.

\begin{equation}
QE = {1 \over T} \Sigma_{t=1}^T || x(t)  - w(t) || 
\end{equation}

\noindent where $x(t)$ is the input sample at the training t; $w(t)$ is the BMU’s weight vector of sample $x(t)$; $T$ is total of training iterations. 

\begin{figure}
    \centering
    \includegraphics[width=0.5\textwidth]{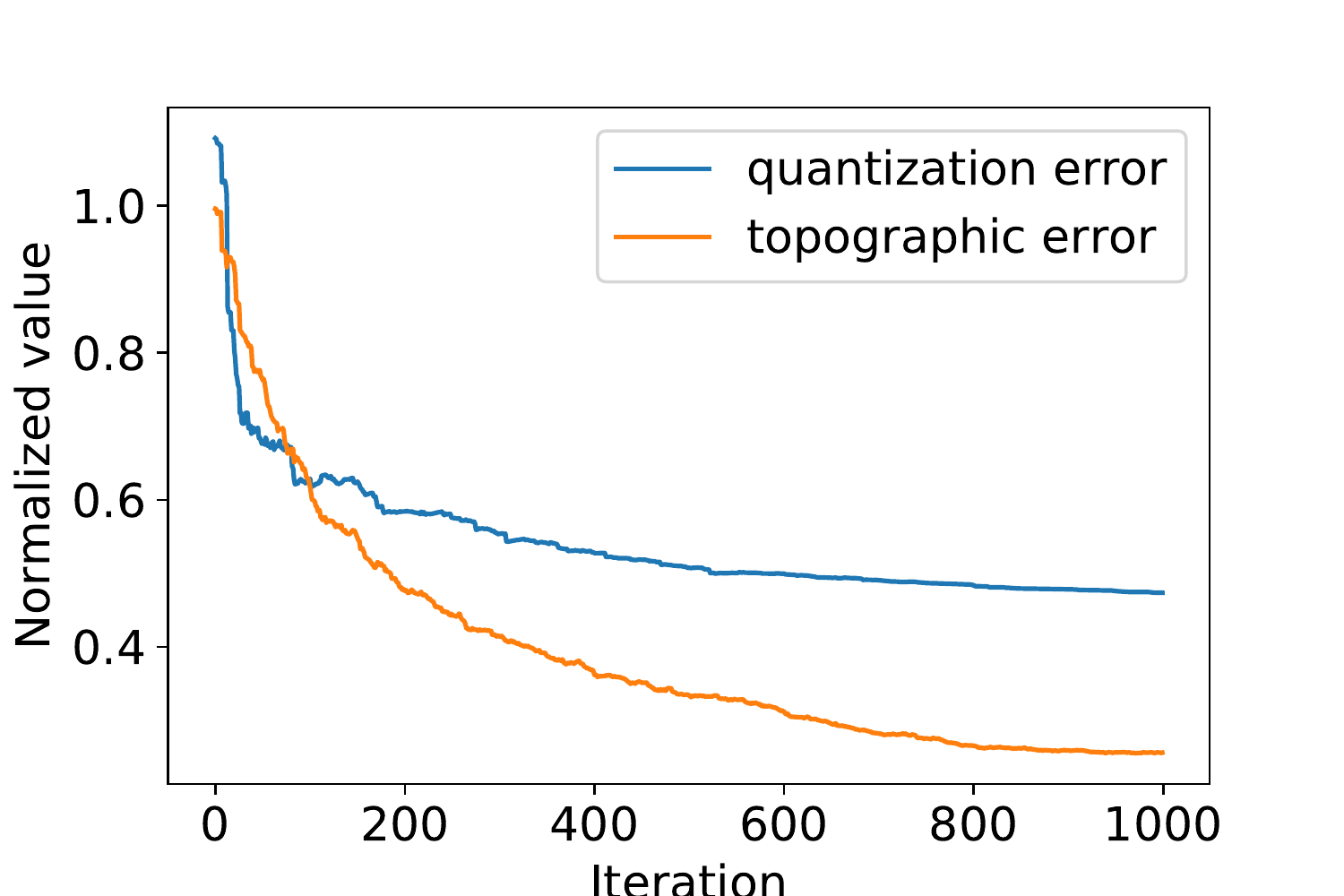}
    \caption{The learning curves of the SOM over 1000 iterations. The Quantization Error (blue) improves first and gain is steady with each iteration. The topographical error (orange) improves a little later (SOM rearranges itself) and then converges.}
    \label{f:som:learning}
\end{figure}

The topographical error indicates the number of the data samples having the first best matching SOM node (SOM-1) and the second best matching SOM node (SOM-2) being not adjacent, i.e. how well is the sample segregated and grouped together in similarity? In a well assembled SOM, the closest node and the next-to-closest node are expected to be adjacent, and therefore this fraction should be small. 

\begin{equation}
TE = {1 \over T} \Sigma_{t=1}^T d(x(t)), 
\end{equation}

\noindent where $x(t)$ is the input sample at training times $t$; $d(x)$ is a step function where, $d(x(t)) = 1$ if SOM-1 and SOM-2 (the closest and second closest match in the SOM) for $x(t)$ are \textit{not} adjacent and $d(x(t)) = 0$ if they are. T is total of all training times. The more often the two best SOM nodes are adjacent, the lower TE will be. Figure \ref{f:som:learning} shows the two learning curves for 1000 iterations. Both stabilize after 1000 iterations and we adopt (and save) this SOM for further use. 

\section{Results}
\label{s:results}
We explore three sets of properties as they are mapped onto the SOM. First the classification by the K-means clustering algorithm. Secondly, the position of green valley galaxies as mapped onto the SOM. And thirdly, we explore how GalaxyZoo voting is distributed over the SOM.

\subsection{K-Means Clusters}
\label{s:som:kmeans}

\begin{figure*}
    \centering
    \includegraphics[width=0.49\textwidth]{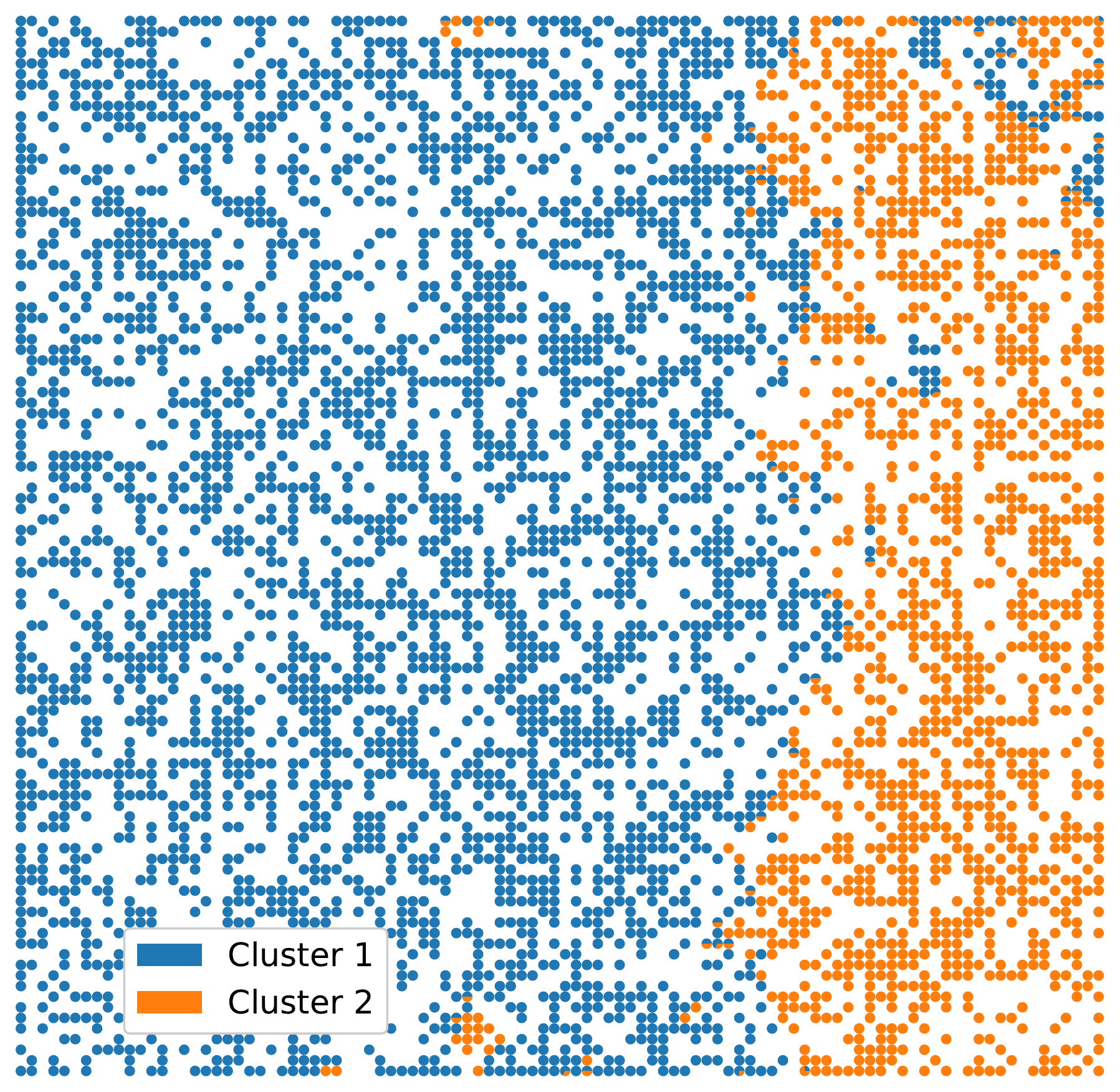}
    \includegraphics[width=0.49\textwidth]{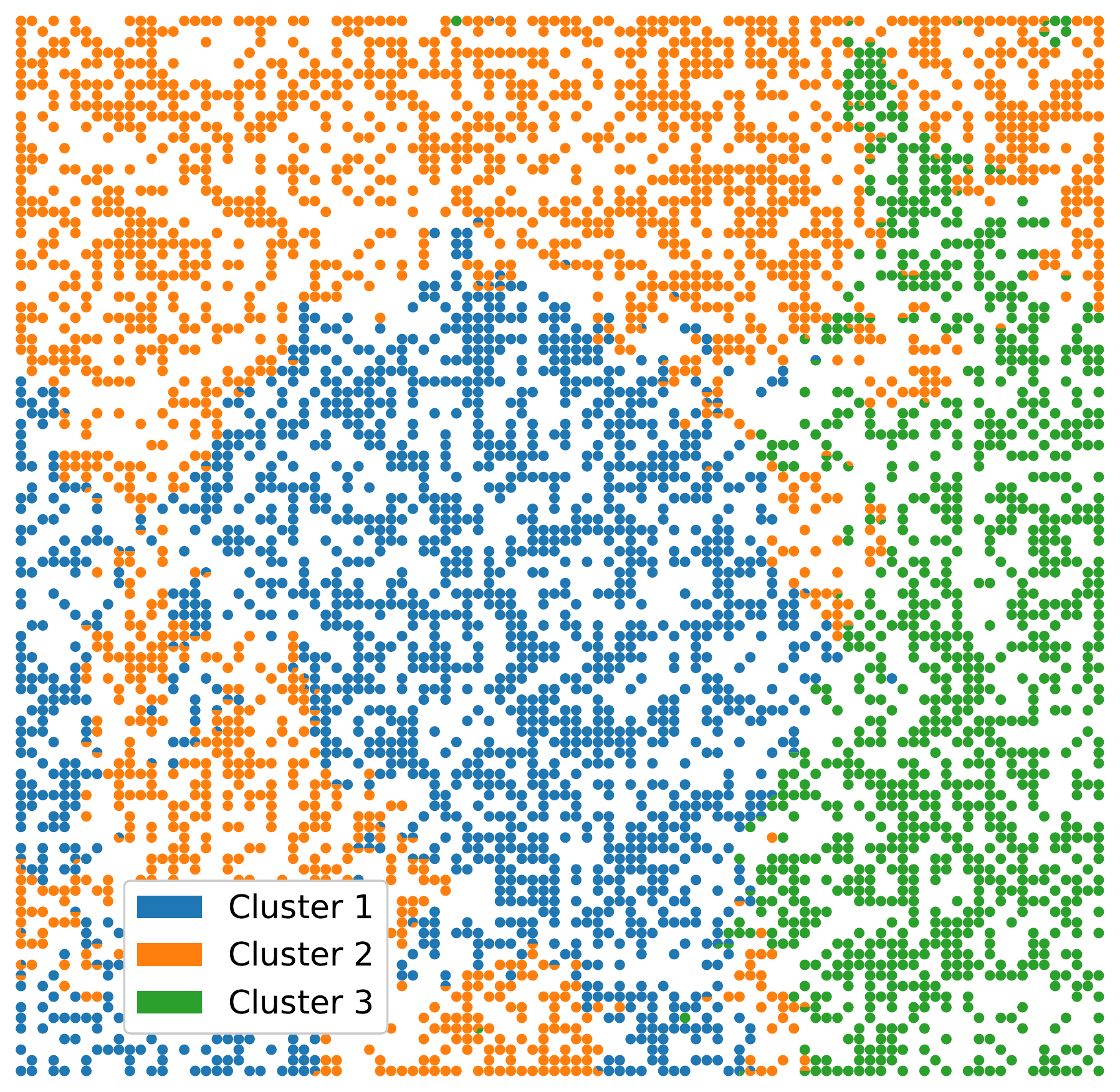}\\
    \includegraphics[width=0.49\textwidth]{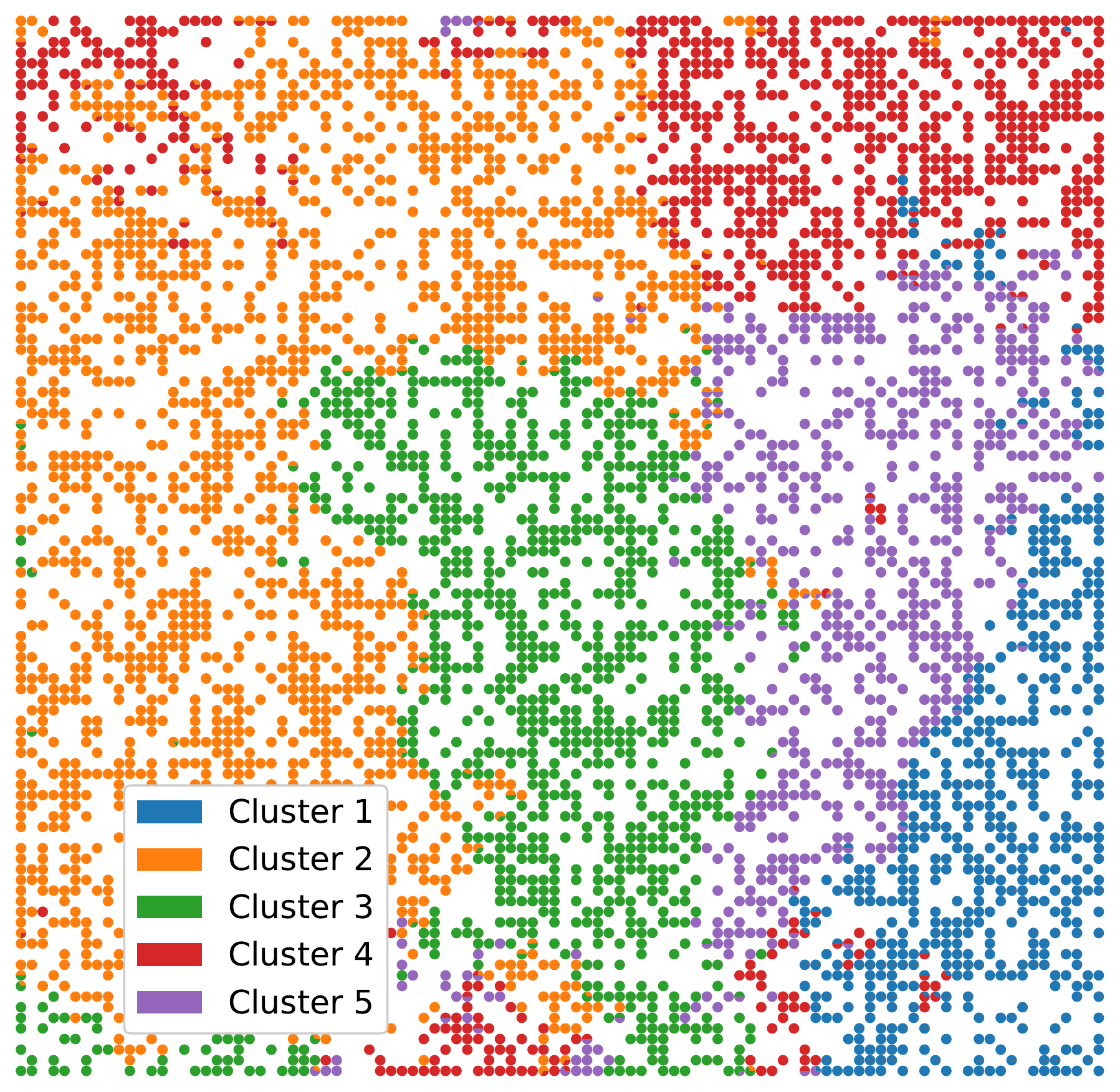}
    \includegraphics[width=0.49\textwidth]{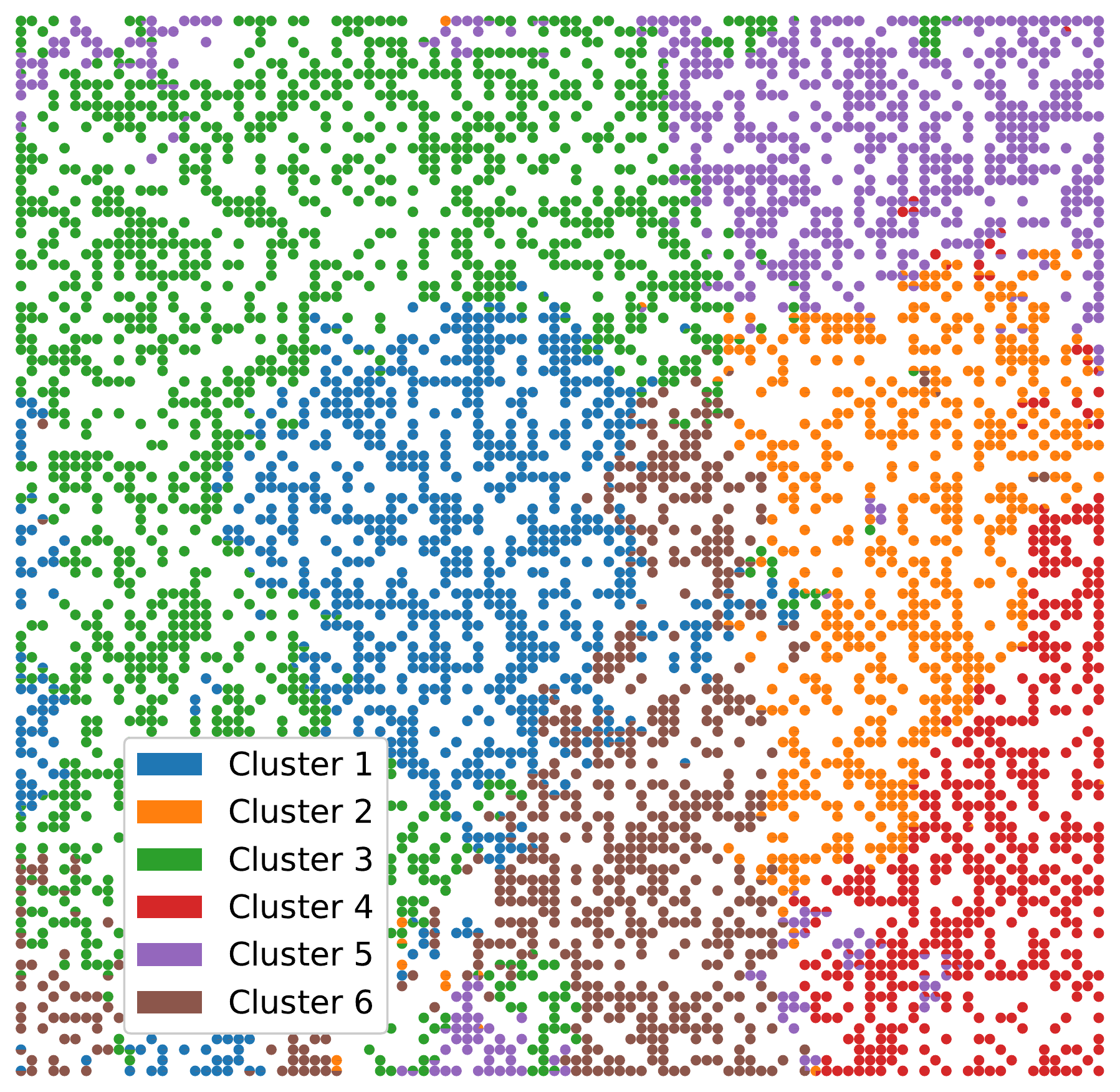}
    \caption{Pie diagrams of each cluster of the $100\times100$ SOM trained on the GAMA feature set. The pie diagrams show if unique K-means clusters are attributed to each SOM node.  
    The amount of mixing is an index of the quality of the clustering. For the most part, K-means clusters are mapped onto unique SOM nodes with little mixing. The K2 and K3 clustering remain coherent (mostly single continuous areas on the SOM). 
    }
    \label{f:som:kclusters:pies}
\end{figure*}

Our first objective is to evaluate how the different K-means clusters are mapped onto the SOM. Figure \ref{f:som:kclusters:pies} shows how each of the clusters from \cite{Turner19} is distributed onto the SOM. In each of the mappings of the K-means clusters onto the SOM, very few nodes are split between multiple K-means clusters showing that the SOM discriminates well between these in the feature space (the SOM is both big enough and well-trained enough). The feature space has enough resolution to make the separation meaningful for the K-means clusters.

From Figure \ref{f:som:kclusters:pies} it appears that K2 or K3 works slightly better on this feature space than K5 of K6. The low degree of mixing between object in different clusters suggests that K2 or K3 clusters provide a better representation of our sample, compared to K5 or K6 clusters. The latter two break into too many sub groupings to improve classification much over K2 or K3. 

\subsection{Green Valley Galaxies}
\label{s:som:gv}

A prime example of an intermediate population of galaxies is the ``green valley", the galaxies in global colour that sit between the star-forming sequence and the quiescent red cloud. Green valley galaxies' intermediate colours are commonly interpreted as a population of galaxies transitioning from star-forming to passive. 

\begin{figure*}
    \centering
    \includegraphics[width=0.49\textwidth]{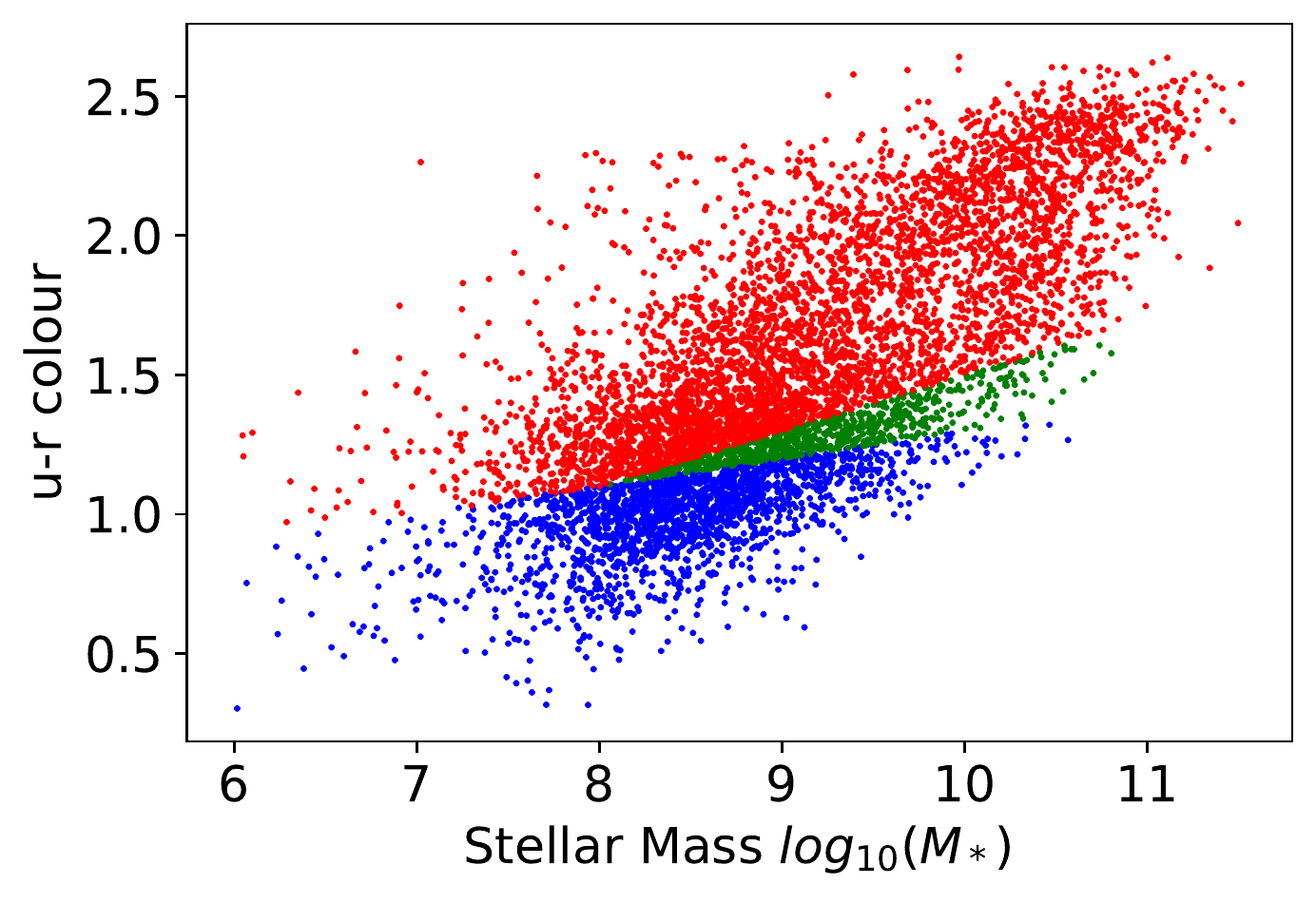}
    \includegraphics[width=0.49\textwidth]{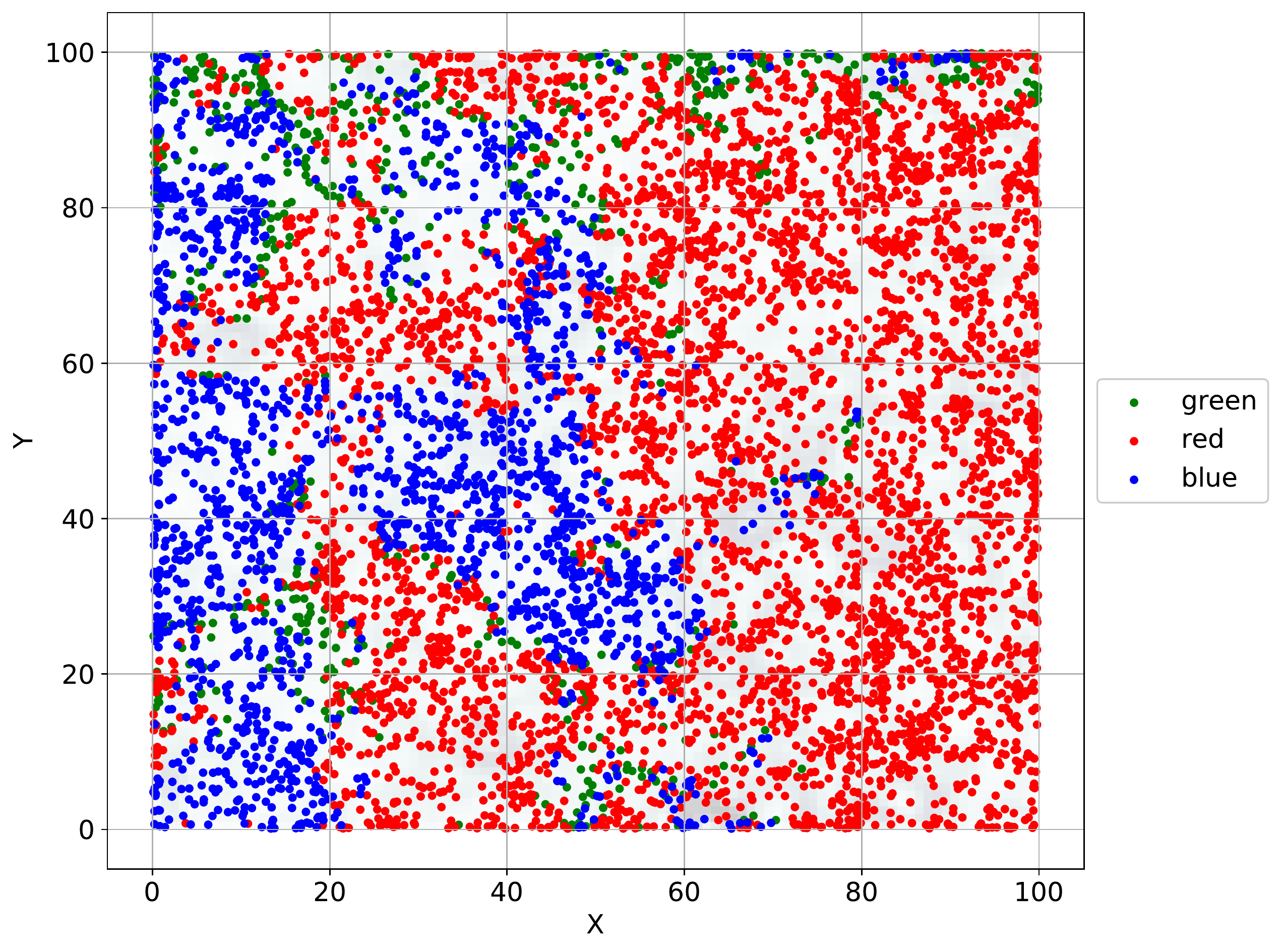}
    \caption{The $u-r$ colour as a function of stellar mass for the GAMA sample of \protect\cite{Turner19}. The green valley criteria from \protect\cite{Bremer18} are shown for the full mass range. 
    The position of the green valley galaxies identified in Figure \ref{f:greenvalley} on the SOM. Green valley galaxies, a canonical transitional population in one part of this feature space, are found all over the SOM map of the full feature space.
    }
    \label{f:greenvalley}
\end{figure*}

\cite{Bremer18} present a working definition based on stellar mass and the restframe, dust corrected $u-r$ colour, also adopted in \cite{Kelvin18}. We start with the adoption of this colour criterion for the green valley but extrapolate the definition to the entire mass range of our sample rather than the narrow mass range used in \cite{Bremer18}. 
Figure \ref{f:greenvalley} shows the colour cuts in $u-r$ colour --uncorrected for dust-- and stellar mass space. The green valley population starts a little above $10^9 M_\odot$ stellar mass and the division at lower mass is just between red and blue galaxies. This is consistent with the picture in \cite{Taylor11} who model the stellar mass and $u-r$ colours of GAMA galaxies.

Figure \ref{f:greenvalley} shows where the red and blue galaxies and the green valley galaxies according to the uncorrected \cite{Bremer18} colour criterion are mapped onto the SOM. Both red and blue galaxies are made up of several clusters on the map and green valley galaxies only concentrate on a few nodes in between the red and blue populations. The blue and red populations are spread over the SOM as blue and red subpopulations end up at different nodes, separated e.g. by stellar mass. The blue and red coherence is driven by the colour and specific star-formation rate, with the substructure and partial fragmentation reflecting the more distinct galaxy red and blue sub-populations.


\begin{figure*}
    \centering
    \includegraphics[width=0.49\textwidth]{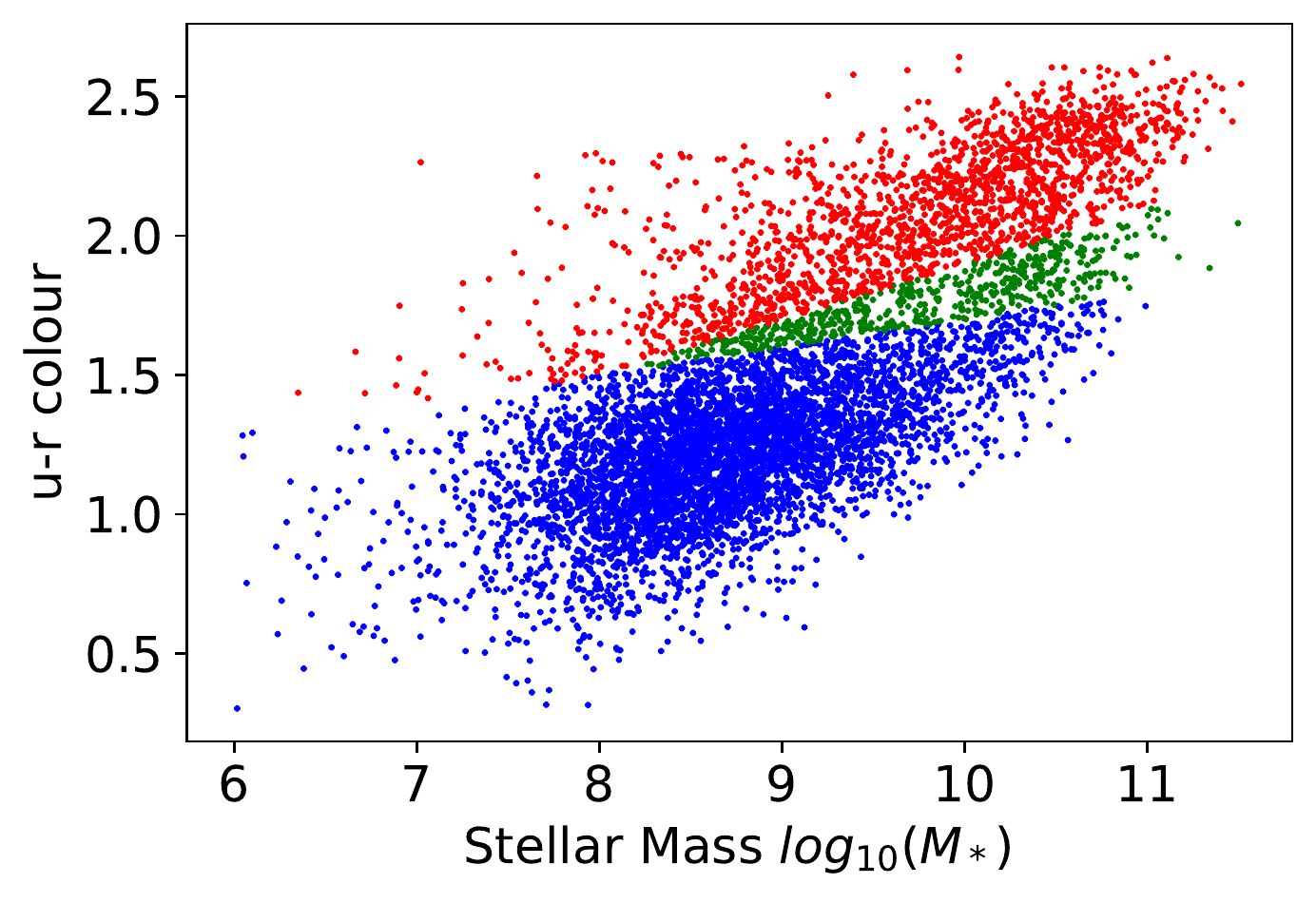}
    \includegraphics[width=0.49\textwidth]{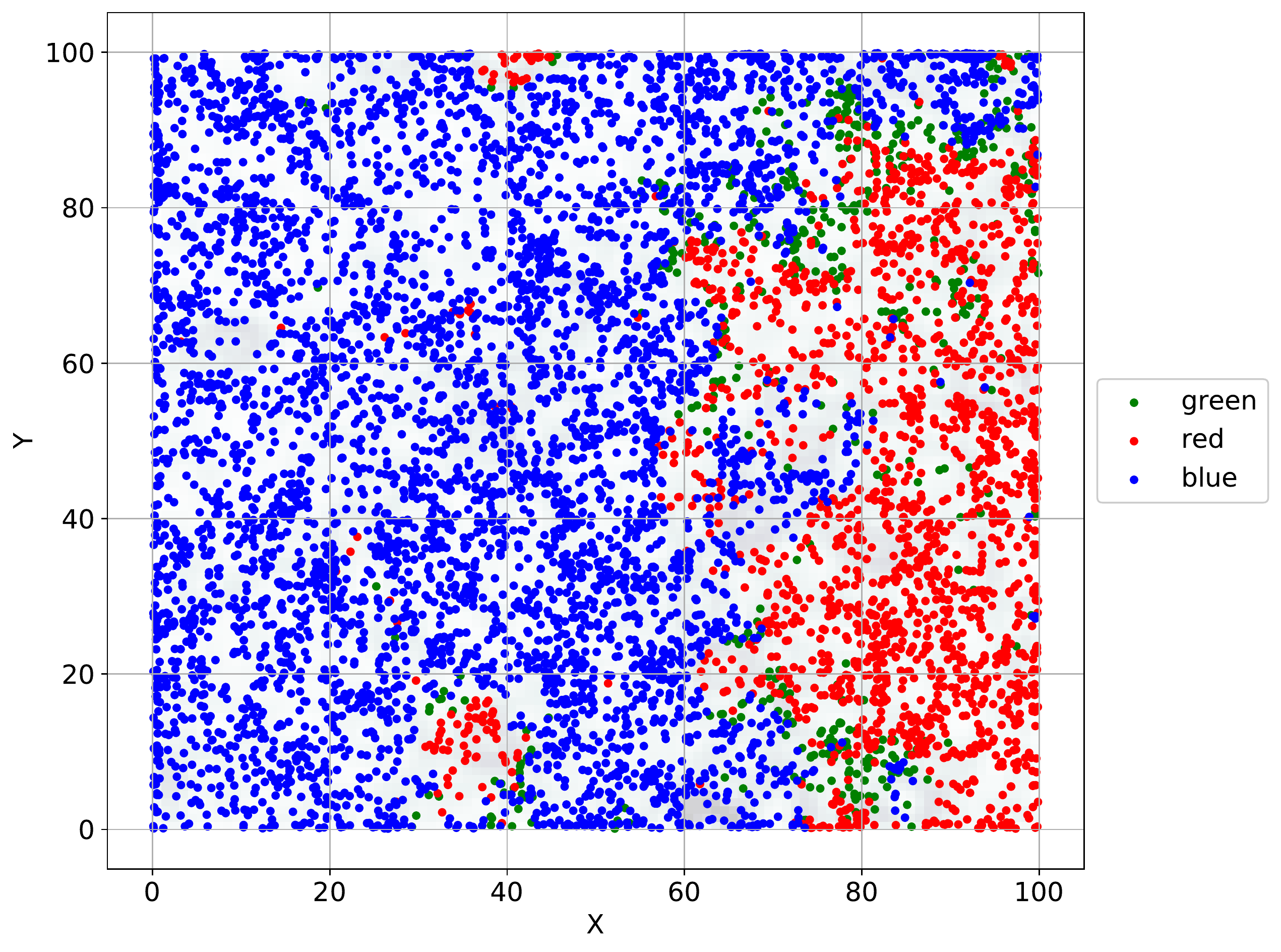}
    \caption{Left: The $u-r$ colour as a function of stellar mass for the GAMA sample of \protect\cite{Turner19} corrected for extinction ($u-r+0.4$) to bring these in line with \protect\cite{Bremer18}. The green valley criteria from \protect\cite{Bremer18} are shown for the full mass range. 
    Right: the position of the green valley galaxies identified on the SOM. Green valley galaxies, a canonical transitional population in one part of this feature space, are found all over the SOM map of the full feature space. Compare this colour corrected for dust (as a population, not individually) with the specific star-formation criterion in Figure \ref{f:greenvalley:salim15}.}
    \label{f:greenvalley-cor}
\end{figure*}

\cite{Turner19} used $u-r$ colours that had not been corrected for internal dust attenuation and this has been our feature space. However, \cite{Bremer18} use the $u-r$ colours that are dust-corrected. Here, we apply an general offset to the \cite{Bremer18} green valley definition in order to reflect this difference. \cite{Turner19} did something similar to divide between red and blue galaxies. Following this, we apply a +0.4 mag shift to the $u-r$ colour criteria from \cite{Bremer18} in Figure \ref{f:greenvalley-cor} and show where the corrected colours land on the SOM. 

If we compare Figure \ref{f:som:featuremap} to Figure \ref{f:greenvalley-cor}, several sub-populations can be identified. Previously, \cite{Schawinski14} pointed out that the green valley is no single population of galaxies but a mix of several intermediate groups \citep[see also][]{Smethurst15,Moutard16}. 

There are two red/green clumps amidst the blue populations at x=35,y=10 and x=40,y=100 on the SOM visible in Figure \ref{f:greenvalley-cor}. These correspond to different colours and high-mass galaxies in Figure \ref{f:som:featuremap}. Concentrations of green galaxies can be identified around X=80,Y=10 and X=80,Y=90, correspinding to high-mass and extended galaxies with a steep profile ($log(n) \simeq 1$) characteristic of ellipticals or bulge-dominated disk galaxies.

The SOM applications show that in the feature space, the green valley galaxies are not one single intermediate group (i.e. a single group of nodes with just green galaxies) but several small intermediate populations combined. 
Morphology studies of the green valley have shown that it is made up of both disk galaxies and ellipticals, and the galaxy disks are dimming and perhaps forming rings \citep[][Smith et al. {\em in prep.}]{Kelvin18,Fernandez21}, perhaps driven by bars in secular evolution \citep[e.g,][]{Geron21}.
Morphological features not included in our feature space (e.g. rings and bars) are drivers of this evolution of these sub-populations. The different ``interstitial'' sub-populations are distinct enough in the galaxy-wide feature space to be separated out in our SOM, but other features may better distinguish them from each other.

The multiple green valley populations sandwiched between red and blue populations on the SOM support the idea  that the green valley population of galaxies is ``interstitial''\footnote{Interstitial meaning in-between population or populations, not necessarily transitioning from one space to the other but settled in a niche between principal spaces.} rather than exclusively a single transitioning population of galaxies. We note that \cite{Bremer18} only employed the green valley criterion in a narrow mass range and for corrected $u-r$ colours. If we restrict ourselves to this mass range ($10.25 < log_{10}(M^*) < 10.75$), we are predominantly left with red galaxies in our sample.

\begin{figure*}
    \centering
    \includegraphics[width=0.49\textwidth]{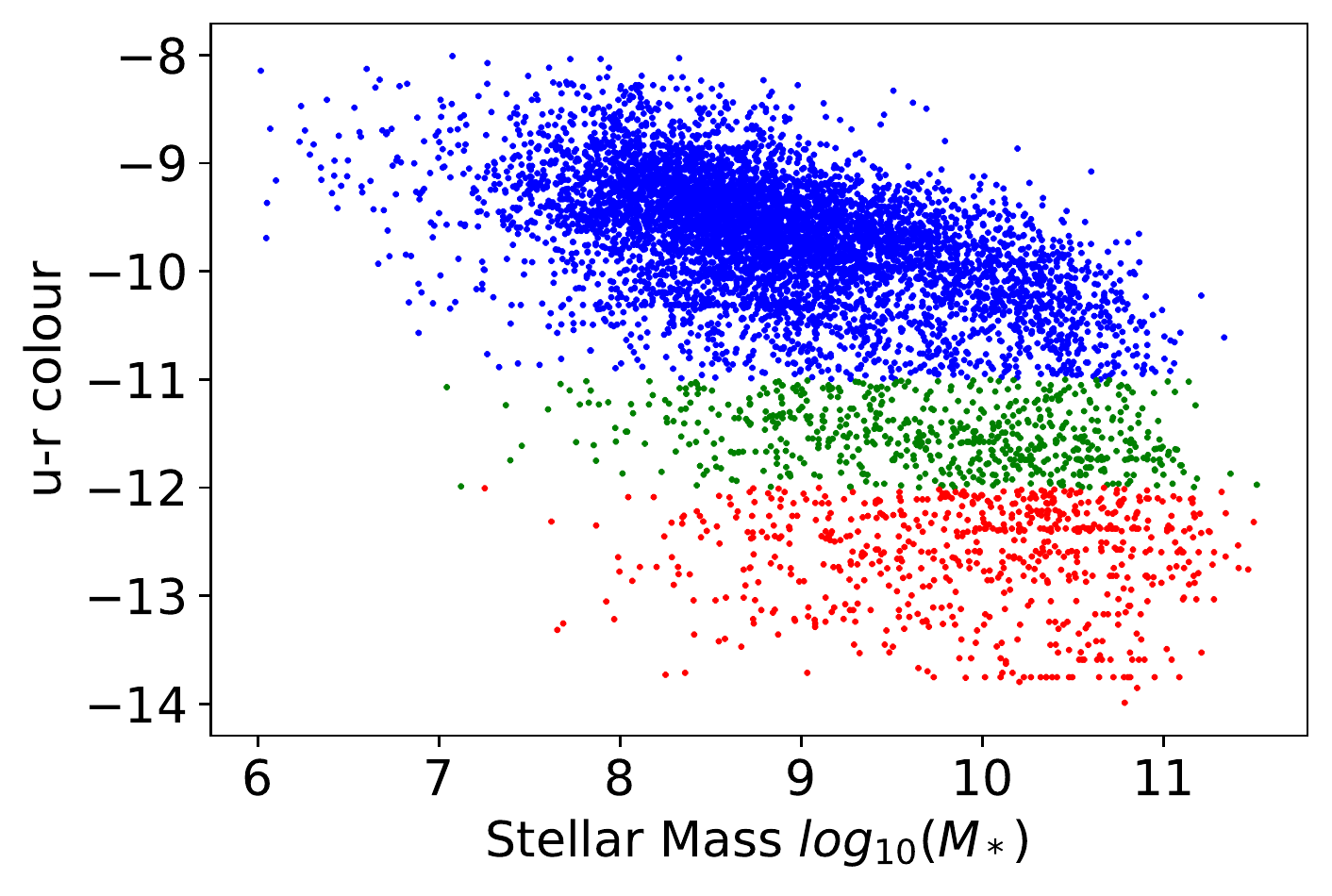}
    \includegraphics[width=0.49\textwidth]{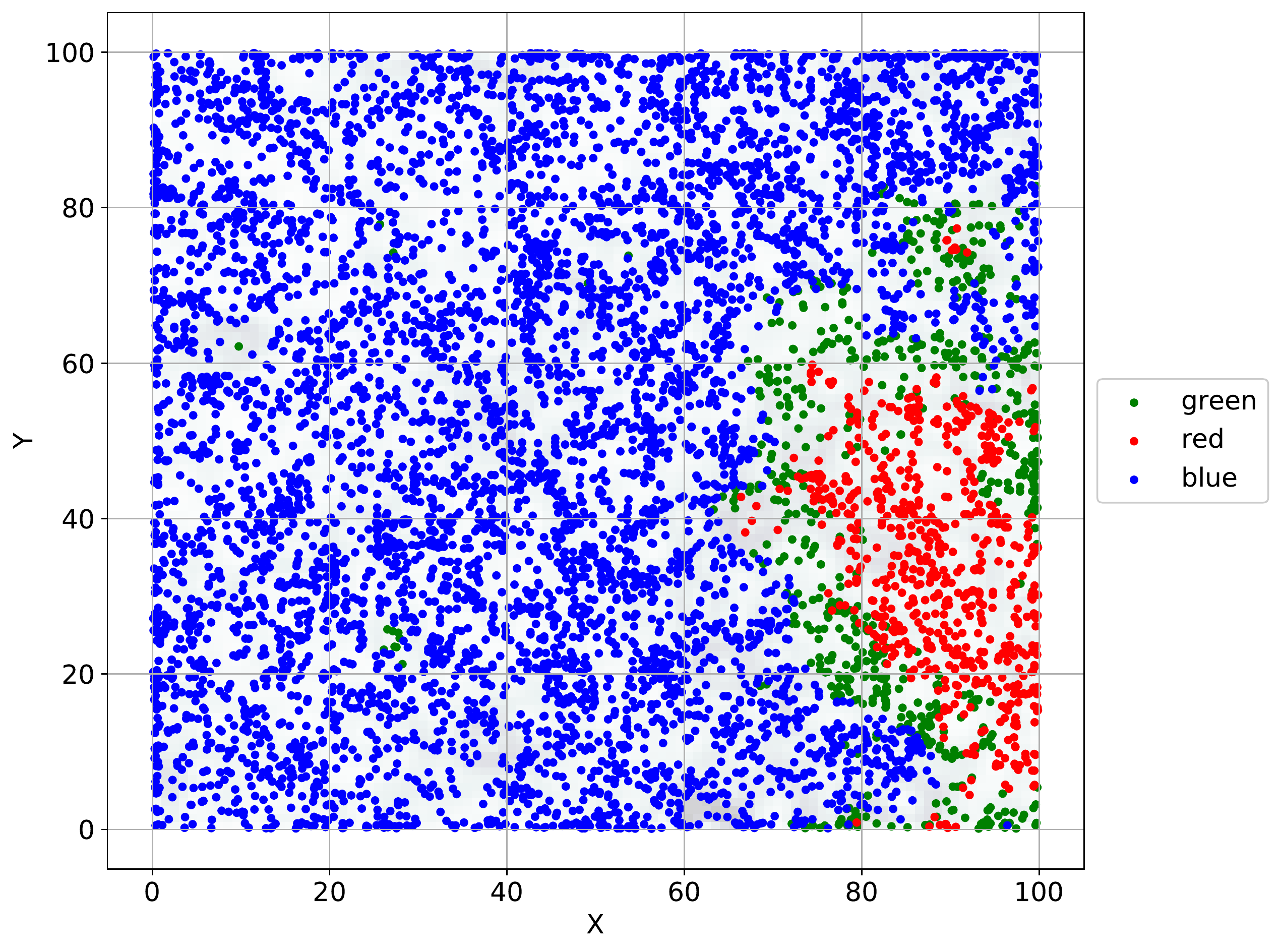}
    \caption{The definition for the green valley using specific star-formation rate from \protect\cite{Salim15} for our nearby sub-sample from \protect\cite{Turner19}. The position of the green valley galaxies according to the criterion in \protect\cite{Salim15} based on the specific star-formation ($-12 < log_10(SSFR) < -11$). This criterion separates the red from the blue populations with a ring of green objects. A few sub-nodes of red galaxies with green borders can be found as well.}
    \label{f:greenvalley:salim15}
\end{figure*}

\cite{Salim15} notes in their overview of green valley galaxy properties that these u-r optical colours may only allow for a poor separation in the level of star-formation. Ultimately, the green valley is defined as an intermediate star-forming population. Figure \ref{f:greenvalley:salim15} shows the green valley as defined as specific star-formation levels between $-12 < log10(SSFR) < -11$. We note that both the uncorrected $u-r$ colour and the specific star-formation were inputs in the SOM training. 

Figure \ref{f:greenvalley:salim15} shows the distribution of blue, green, and red galaxies based on the definition in specific star-formation rate in Figure \ref{f:greenvalley:salim15}. The blue and red populations separate out in two unique groups with the green valley galaxies in between. We note that there are several small ``pockets" of red galaxies unique in S\'ersic index or stellar mass (see Figure \ref{f:som:featuremap}). 

Both the corrected $u-r$ colour definition and the specific star-formation definition of the green valley mapped onto the SOM point to sub-populations in the green valley based on stellar mass, profile, or both.

The K-means clustering was applied by \cite{Turner19} to identify the number of bi-modalities in this parameter space. The green valley is in the saddle point of one of these bi-modalities. We shall briefly compare the position of the green valley on our SOM and compare it to the disposition of the K-means clusters. 

Using the $u-r$ colour definition, we compare Figure \ref{f:greenvalley} to the earlier mapping of the K-means nodes in Figure \ref{f:som:kclusters:pies}. The red and blue clouds correspond fairly closely to different K-means nodes; for example, the red cloud corresponds to cluster 3 in the 3-cluster classification with the blue cloud corresponding mostly to cluster 1 and 2. The green valley galaxies are evenly split as either one or the other K-cluster. 

The corrected $u-r$ colour selected green valley galaxies reside in more specific nodes in the higher K-means classifications. For example, in the 5-cluster classification, the low-mass green valley galaxies are in cluster 4. And in the K-6 cluster solution, the green valley galaxies correspond largely to outlier sections of one of the K-means clusters (cluster 5). Most of the K-means cluster 5 is still a coherent structure on the SOM but the transitioning or interstitial populations are classified in completely different nodes of the SOM. 

Using the specific star-formation definition of the green valley (Figure \ref{f:greenvalley:salim15}), we observe the same trend.
We compare Figure \ref{f:greenvalley:salim15} to the K-means clusters in Figure \ref{f:som:kclusters:pies}, it is clear that red and blue populations are associated with one or two clusters in the K2 or K3 classification. But even in higher order K-means clustering, the green valley galaxies are never grouped in a single K-means cluster and always evenly split between K-nodes. 

\begin{figure*}
    \centering
    \includegraphics[width=0.49\textwidth]{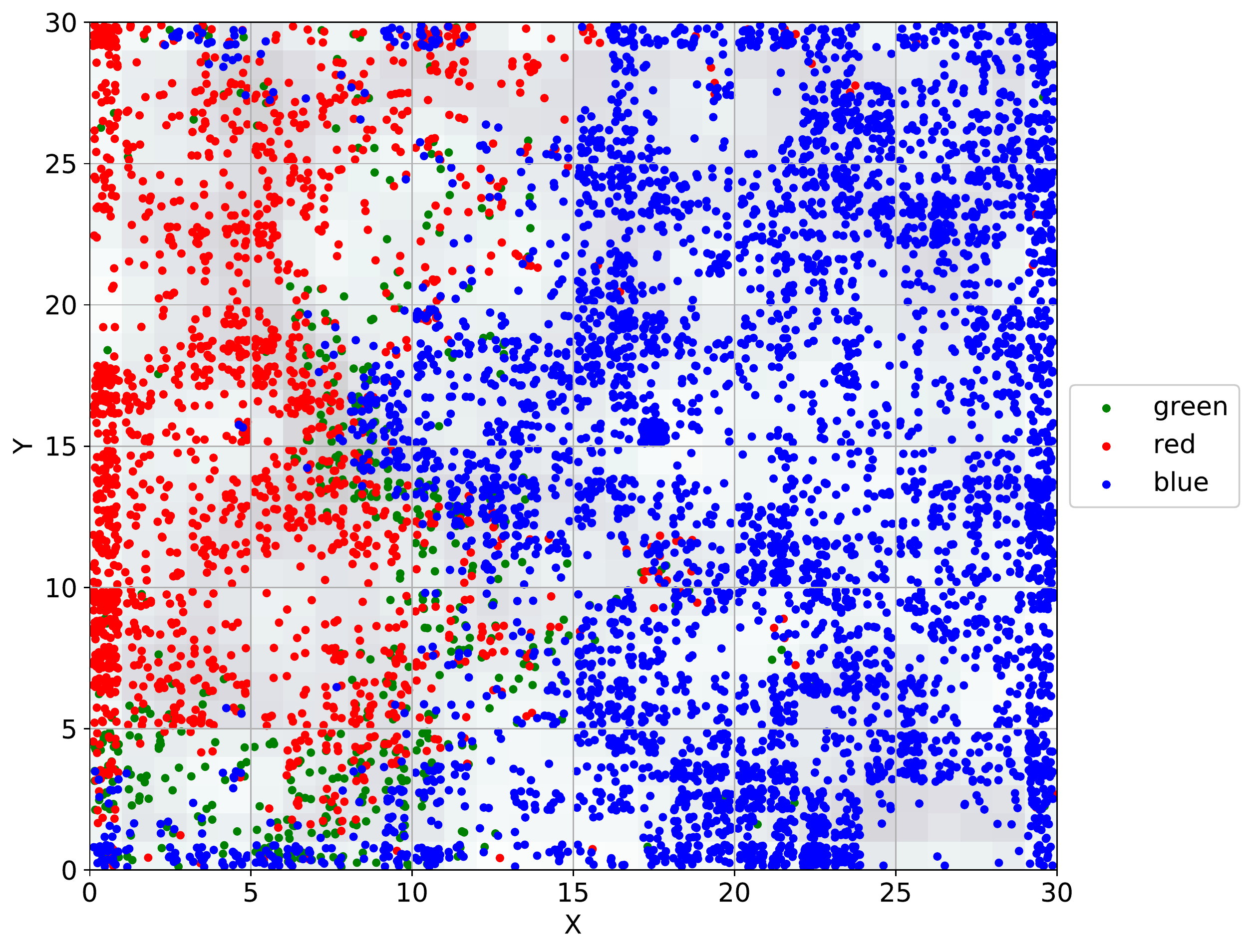}
    \includegraphics[width=0.49\textwidth]{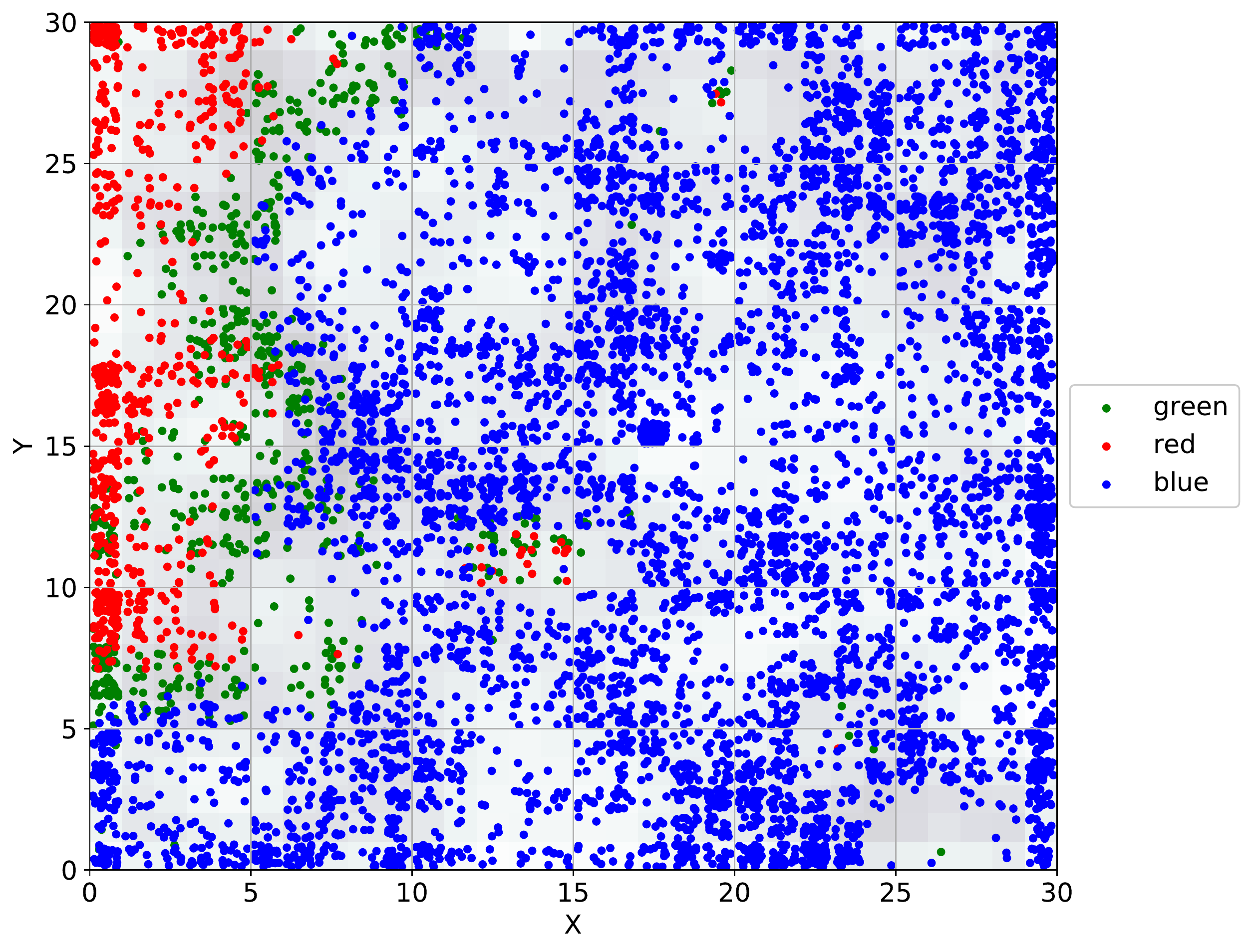}
    \caption{The position of the green valley galaxies identified in the $30\times30$ SOM (see also Figure \ref{f:som:frequency}). Left: the blue, green, and red classification based on corrected $u-r$ colour criteria  shown in Figure \ref{f:greenvalley-cor}. Right: the blue, green, and red classification based on specific star-formation rate shown in Figure \ref{f:greenvalley:salim15}. }
    \label{f:som30:greenvalley}
\end{figure*}

Part of our motivation for this work was to evaluate how good SOM mapping is to identify smaller galaxy sub-populations in a well-understood sample, previously classified by a Machine Learning algorithm. To evaluate the robustness of the green valley conclusions above, we map these populations on a much smaller SOM ($30\times30$) Figure \ref{f:som30:greenvalley} for both definitions of blue, green, and red galaxies. General conclusions still hold. This map cannot be compared directly to Figure \ref{f:som:kclusters:pies} as each iteration of a SOM is unique and depends on starting seed. 
The smaller sub-populations of green valley galaxies is still there but more mixed in with a red sub-population as the resolving power of the smaller SOM mixes these objects together. This argues for slightly ``roomier" SOM choices for galaxy populations if one wants to identify sub-populations of interest.

\subsection{GalaxyZoo Classifications}
\label{s:som:gz}

The GalaxyZoo project \citep{Lintott08} has produced excellent results on galaxy morphology using citizen science voting on specific features. The feature space (Figure \ref{f:gama:features} and \ref{f:som:featuremap}) does not include direct information on morphology in $\sim$kpc scales. The S\'ersic profile's index ($n$) and half-light radius ($r_{50}$) do not contain much information on number of spiral arms or bar fraction directly. One can use the GalaxyZoo classifications of these galaxies \citep[][Kelvin et al., {\em in prep.}]{Holwerda19} as \textit{a posteriori} labels for the SOM. 

Here, we present an example of where voting fractions for several of the disk galaxy questions are mapped onto the SOM. Our aim is to explore how well the feature space of integrated galaxy properties can relate to sub-galaxy scale phenomena such as bulges or bars and the number and winding of spiral arms.

\begin{figure*}
    \centering
    \includegraphics[width=0.49\textwidth]{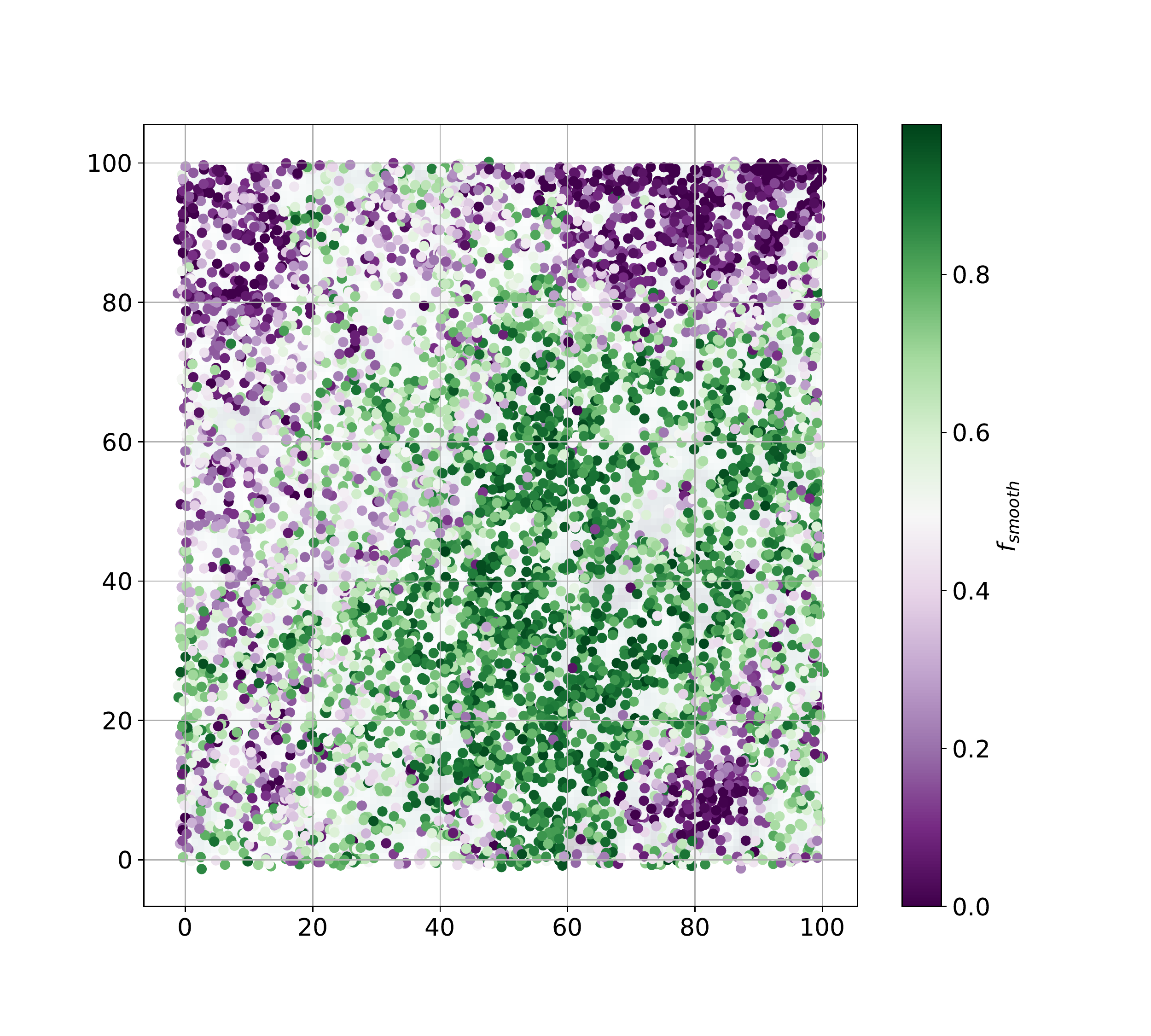}
    \includegraphics[width=0.49\textwidth]{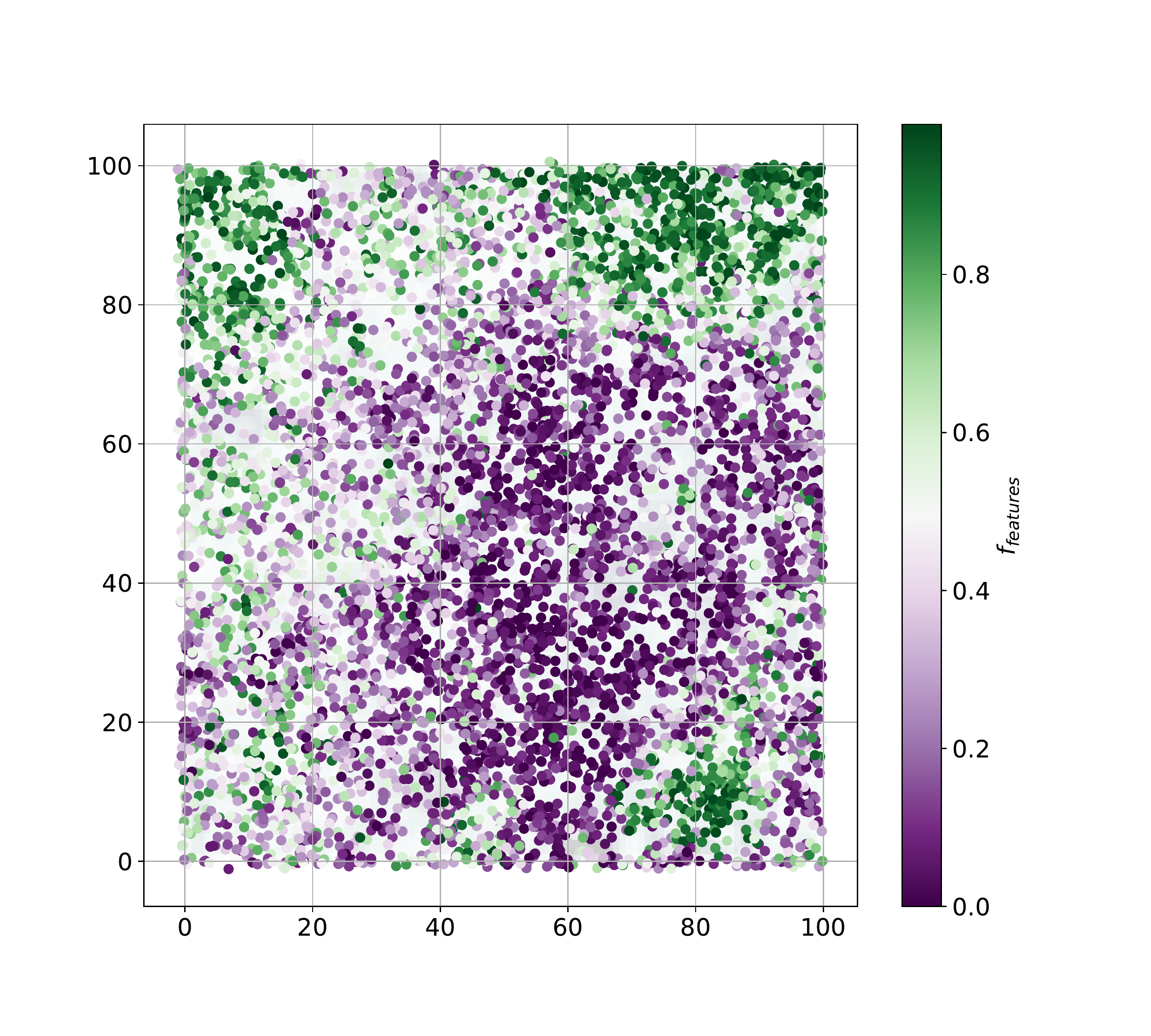}
    \caption{The fraction of GalaxyZoo votes in favor of a smooth galaxy, likely an elliptical, (right), and the voting fraction in favor of a galaxy with ``features", likely a spiral galaxy (left). The votes are complementary identifying a smooth population over the center of the SOM with several features galaxy populations in the corners. The only other choice was ``artifact".}
    \label{f:som:gz:smooth}
\end{figure*}

The very first question in GalaxyZoo is whether the object has features, looks smooth or appears to be a star or artifact. Figure \ref{f:som:gz:smooth} shows the voting fractions in favor of a smooth galaxy and those in favor of a galaxy having ``features".  With S\'ersic profile information as part of the feature space, one would expect a segregation of smooth (elliptical) galaxies and disk (featured) galaxies. The SOM voting in Figure \ref{f:som:gz:smooth} does show a reasonable separation in smooth and ``featured" galaxies\footnote{We note however, that a judicious choice of colour bar was needed in Figure \ref{f:som:gz:smooth} to show this.}.

Our sample is low-redshift and should be well resolved in the KiDS images used for the GalaxyZoo. Other, smaller features, such as whether a galaxy has spiral arm structure or a bar, do not show as much separation on the $100\times100$ SOM. The feature space of stellar mass ($M_*$), colour (u-r), specific star-formation rate (SSFR), S\'ersic index ($n$), and half-light radius ($r_{50}$) does not discriminate well to cleanly isolate such features. 
A different feature space is needed for the separation of sub-galaxy scale morphological structures.

\section{Concluding Remarks}
\label{s:concl}

We have mapped the feature space of a sample of nearly 8000 GAMA galaxies from \cite{Turner19} onto a Self-Organizing Map to explore the success of K-means clustering on a galaxy sample, the green valley of galaxies, and some morphological features identified by the Galaxy Zoo. We opted for a 100$\times$100 node SOM. This size SOM was sufficient to separate out populations in this sample and the feature space contained enough resolution to do so (see Figure \ref{f:som:kclusters:pies}, \ref{f:som:frequency}, and \ref{f:som30:greenvalley}).

Our first result is to compare the K-means clustering and Self-Organizing Maps, which are both instances of unsupervised learning on the entire dataset without necessarily splitting it into training and test or validation samples. The SOM application is a relatively straightforward comparison with the existing classifications from \cite{Turner19}. From their mapping onto our SOM, the K-means clustering using three (K3) or five (K5) clusters seem good descriptions, resolving into clear and continuous regions on the SOM (Figure \ref{f:som:kclusters:pies}). A higher number of clusters (K6) appear to be somewhat of an over-fit. However, the evaluation is still based on the visual interpretation of the SOM and remains a subjective one.

We note here that the GAMA data is not ideal K-means clustering data with an unequal distribution distance between peaks and clear elongation rather than isomorphic clusters. The result from \cite{Turner19} stands: one needs more than a single bimodality in the galaxy population, manifesting in each feature (k2), to explain the bimodalities seen in single feature distributions. This is reflected in their Silhouette Coefficients as well (Table \ref{t:kmeans:silouette}).

Secondly, we use two definitions of a traditional classification in a part of our feature space: the division into red and blue galaxies with the green valley in between based on $u-r$ colour and on specific star-formation. 
Green valley galaxies are indeed an interstitial population but not a single coherent one; several small green sub-populations are scattered throughout the SOM. Even using the general galaxy-wide properties the green valley population is spread to several different parts of the map (Figures \ref{f:greenvalley} and \ref{f:greenvalley:salim15}).
Blue and red galaxies separate based on the feature space into a single almost coherent grouping (albeit a complex shape) on the SOM. The green valley galaxies are a few nodes between the two dominating populations. We argue that this is consistent with the green valley being made up of several interstitial sub-populations, each quite distinct. Some may indeed be transitioning from blue to red (or the reverse) and some are not \citep[consistent with ][]{Taylor11,Schawinski14}.
Especially notable is that low-mass green valley galaxies are mixed in with the other lower mass galaxies (Figure \ref{f:greenvalley-cor}) consistent with these being much more alike then they are at higher masses. 

Our final SOM experiment is to evaluate if the galaxy-wide feature space ($M_*$, $SSFR$, $u-r$, S\'ersic $n$ and $r_{50}$) can map morphological features, such as those expressed by GalaxyZoo labels.

As a proof of concept, the first GalaxyZoo question (smooth or featured?) voting pattern for these galaxies does appear to segregate well on the SOM. Further GalaxyZoo questions hardly show any separation (e.g. the presence of a bar or spiral structure) which dominate in featured galaxies. Detailed morphologies on the kpc scale (bulges, spiral arms, and bars) are not well separated by this SOM map using the general galaxy properties feature space ($M_*$, $SSFR$, $u-r$, S\'ersic $n$ and $r_{50}$). A different feature space is needed to classify to this detail.

A Self Organizing Map looks to be a promising tool to identify known and unknown populations in galaxy catalog feature space. The sub-populations in the green valley are our first example of that. We note that a slightly wider SOM than recommended aids with the separation of such sub-groupings. The hope for future applications is to identify instructive sub-populations in galaxy surveys using this mapping technique.

\subsection{Data availability}

We use the data-tables of the feature space and the K-means labels from \cite{Turner19}, derived from the GAMA DR3 \citep[][\url{http://www.gama-survey.org/dr3/}]{Baldry18} DMUs \citep[{\sc SersicCatSDSS}][]{Kelvin12}, ({\sc MagPhys}, S. Driver), based on {\sc LAMBDAR} photometry \citep[{\sc LambdarPhotometry}][]{Wright16}, and the {\sc minisom } \citep{minisom} package and examples. 

\section{Acknowledgements}

BWH is supported by an Enhanced Mini-Grant (EMG). The material is based upon work supported by NASA Kentucky under NASA award No: 80NSSC20M0047.

AHW is supported by an European Research Council Consolidator Grant (No. 770935).
MB is supported by the Polish National Science Center through grants no. 2020/38/E/ST9/00395, 2018/30/E/ST9/00698 and 2018/31/G/ST9/03388, and by the Polish Ministry of Science and Higher Education through grant DIR/WK/2018/12.

This research made use of Astropy, a community-developed core Python package for Astronomy \citep{Astropy-Collaboration13,Astropy-Collaboration18}.

\bibliographystyle{mnras}

\clearpage
\newpage
\appendix

\begin{figure*}
\section{Feature Space mapped onto the SOM}
    \centering
    \includegraphics[width=0.49\textwidth]{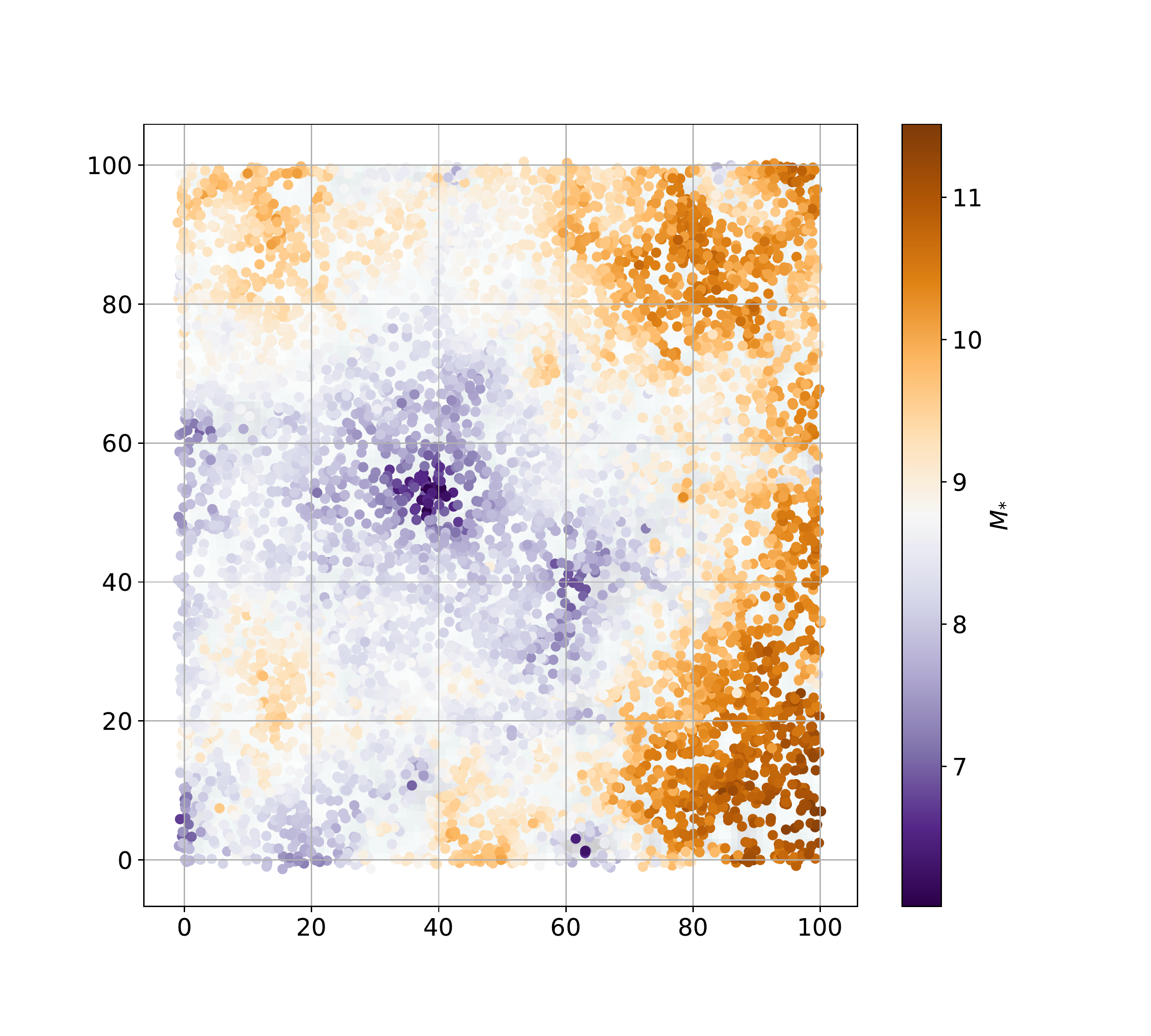}
    \includegraphics[width=0.49\textwidth]{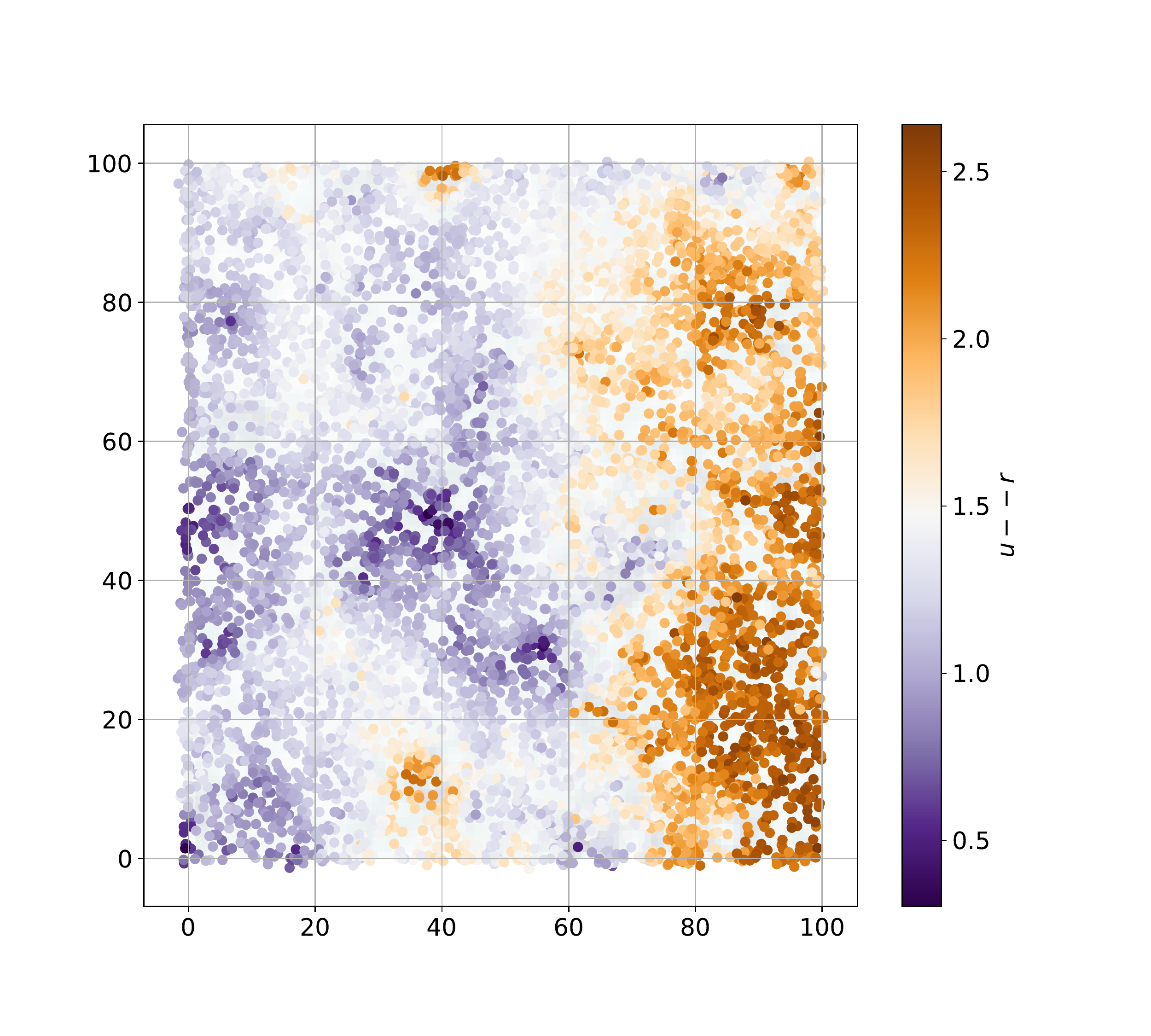}
    \includegraphics[width=0.49\textwidth]{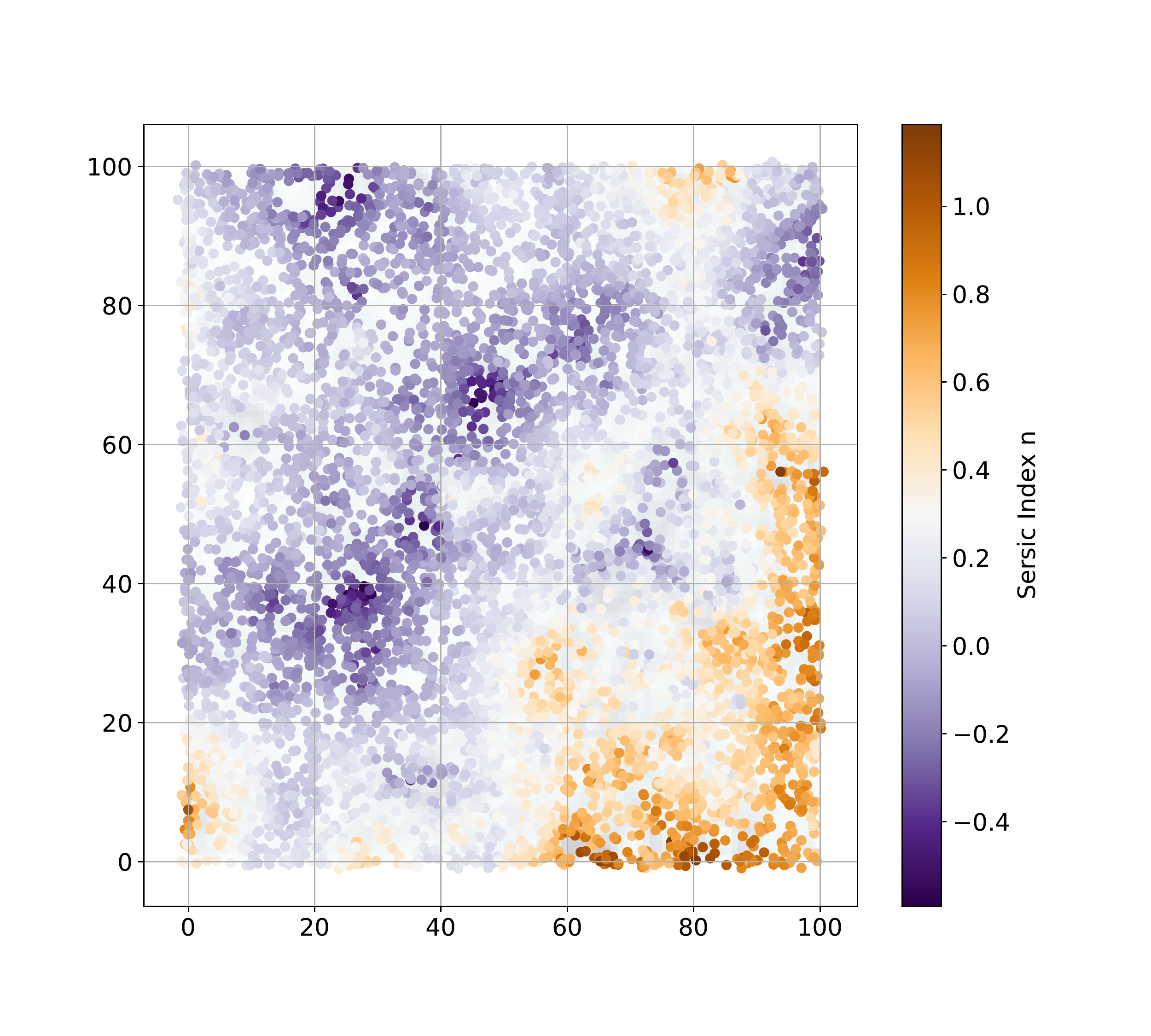}
    \includegraphics[width=0.49\textwidth]{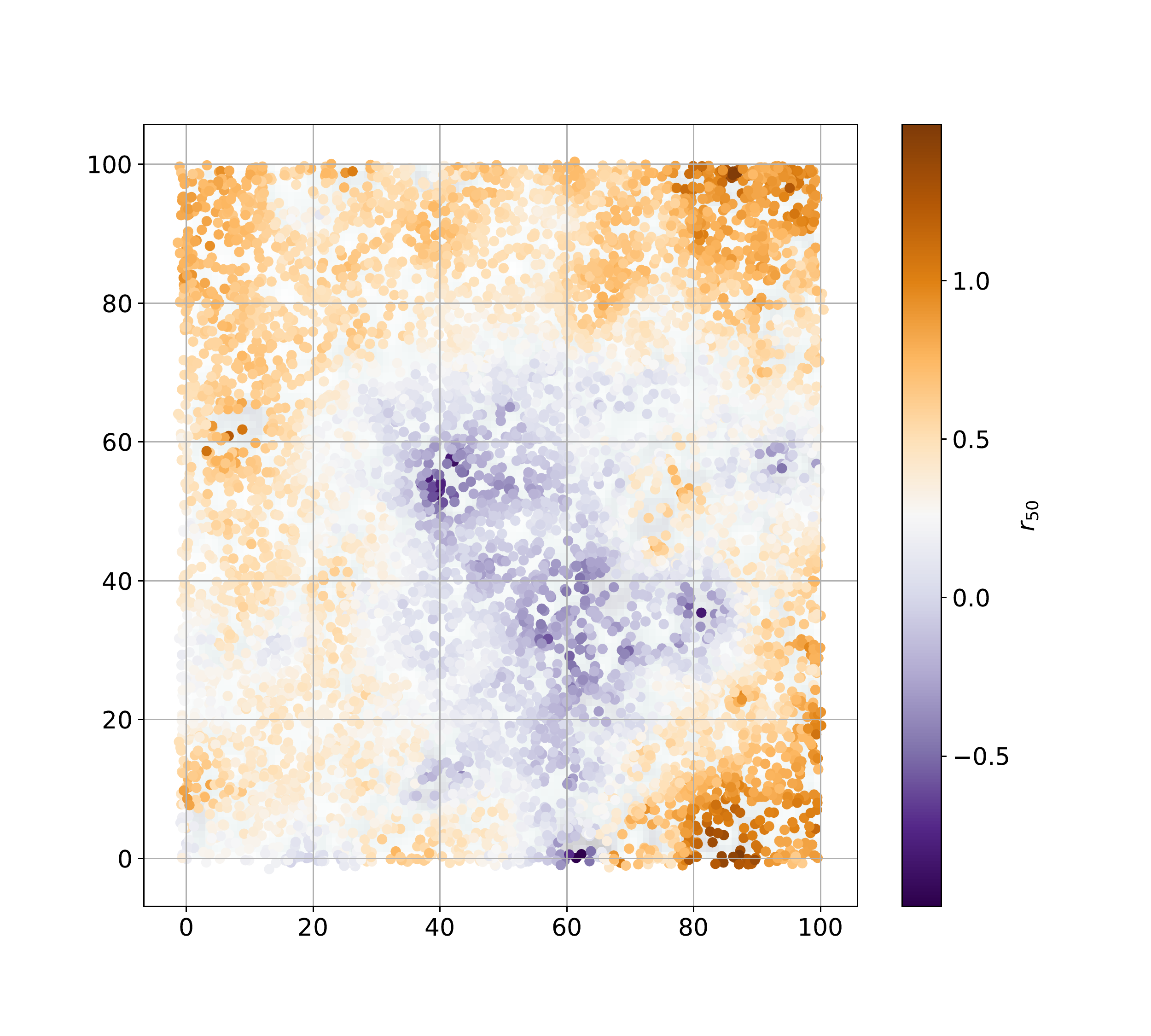}
    \includegraphics[width=0.49\textwidth]{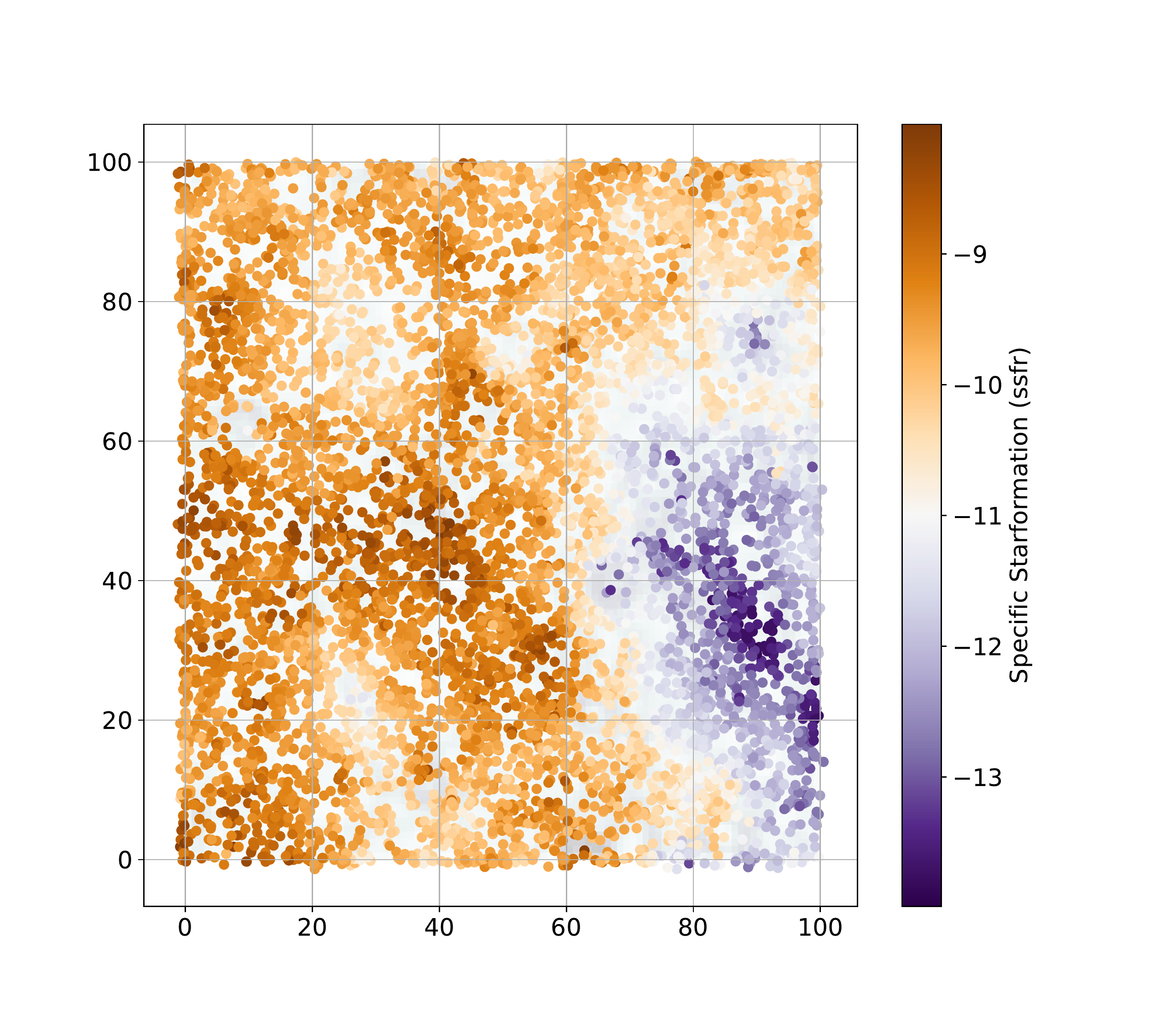}
    \caption{The absolute values of the feature space as mapped on the SOM. Compare to the weighting in Figure \ref{f:som:featuremap}.}
\end{figure*}

\bsp	
\label{lastpage}
\end{document}